\documentclass[reprint,amsmath,amssymb, aps]{revtex4-2}
\usepackage{graphicx}
\usepackage{dcolumn}
\usepackage{bm}
\usepackage{hyperref}
\usepackage{xcolor}
\usepackage{caption}
\usepackage{subcaption}
\usepackage[linesnumbered,ruled]{algorithm2e}

\begin{document}

\title{Composition-Weighted Symbolic Regression for General-Purpose Property Prediction}

\author{Yang Huang}
\affiliation{
  University of Science and Technology of China, Hefei 230026, China
}
\affiliation{
  Suzhou Institute for Advanced Research, University of Science and Technology of China, Suzhou 215213, China
}
\author{Jingrun Chen}
\affiliation{
  School of Mathematical Sciences and Suzhou Institute for Advanced Research, University of Science and Technology of China
}
\date{\today}

\begin{abstract}
We introduce a composition-weighted symbolic regression framework for interpretable prediction of materials properties directly from chemical composition. The method jointly learns analytical functional forms and task-dependent elemental weightings without predefined descriptors. By incorporating $\max$/$\min$ operators, it naturally enforces constraints such as non-negative band gaps and bounded classification probabilities, unifying regression and classification tasks. Efficient search is achieved through a hybrid Monte Carlo tree search--genetic programming algorithm with gradient-based refinement and parallel computation. Benchmarks on MatBench tasks show competitive accuracy relative to state-of-the-art black-box models while yielding explicit analytical expressions. Applied to III--V semiconductor alloys, the model produces smooth composition-dependent trends and learned elemental weights with chemically meaningful periodic behavior. This framework provides a scalable and interpretable route for materials discovery and property screening.
\end{abstract}
\maketitle

\section{Introduction}

Machine learning for materials property prediction is commonly divided into structure-based and composition-based approaches. Structure-based models often achieve high accuracy by explicitly exploiting atomic configurations and local environments~\cite{xie2018,schutt2018,chen2019,gasteiger2020,dunn2020}. Representative examples include interatomic-potential frameworks such as Equiformer~\cite{liao2022}, TACE~\cite{kim2026}, SevenNet~\cite{park2024}, and DPA~\cite{zhang2024,zhang2024pretraining}, as well as structure-informed property-prediction models including MODNet~\cite{de2021a,de2021b}, coGN~\cite{ruff2024}, AMMExpress~\cite{dunn2020}, Finder~\cite{ihalage2022}, and CrabNet~\cite{wang2021}. These approaches, however, depend on reliable structural information, which is frequently unavailable, uncertain, or computationally expensive to obtain.

Composition-based methods instead predict materials properties directly from chemical formulas, enabling rapid screening over vast compositional spaces without structural relaxation or first-principles calculations~\cite{wang2021,wang2022,de2021a,dunn2020}. Despite their efficiency, most current composition-based models rely on neural-network architectures or other black-box learners whose internal representations are difficult to interpret physically. This creates a central challenge for general-purpose materials informatics: how to retain competitive predictive accuracy while recovering transparent and chemically meaningful analytical relationships.

Here we propose a composition-weighted symbolic regression framework that integrates interpretable composition-based modeling with data-driven functional discovery. A material property $P$ is expressed as
\begin{equation}
  \label{eq:general_form}
  P = \mathcal{F}(\bm{x}; \bm{\theta}), \quad 
  x_k = \sum_i w_{k,i} c_i,
\end{equation}
where $c_i$ is the elemental composition fraction and $w_{k,i}$ are learnable elemental weights. The variables $\bm{x}$ represent composition-weighted averages of latent elemental properties, while $\mathcal{F}$ is an analytical function identified via symbolic regression. This formulation can be interpreted as learning effective elemental properties that combine nonlinearly to reproduce macroscopic observables.

Unlike existing composition-weighted models with predefined functional forms, such as linear or hand-crafted nonlinear descriptors~\cite{meredig2014,ward2016,goodall2020,guo2022}, our approach learns both the functional form and elemental representations directly from data. Both the elemental weights $\bm{w}$ and the function parameters $\bm{\theta}$ are optimized during training, enabling physically interpretable yet flexible predictions. By mapping compositions to a low-dimensional space of composition-weighted variables, our method mitigates the combinatorial complexity of symbolic regression and enables its application to general property prediction.

The resulting framework provides a general, interpretable, and scalable approach to property prediction from chemical composition alone, enabling predictions without predefined descriptors or prior physical assumptions.

\section{Method}

The central task is to jointly determine the functional form $\mathcal{F}$ and optimize the associated elemental weights $\bm{w}$ and parameters $\bm{\theta}$ in Eq.~\eqref{eq:general_form}. We address this problem through a hybrid framework that combines symbolic regression for functional discovery with gradient-based optimization for continuous parameter estimation.

\subsection{Operator Set}

The symbolic search space includes standard continuous operators, such as $\exp(\cdot)$, $\log(\cdot)$, multiplication, and addition, together with the non-smooth operators $\max(\cdot,\cdot)$ and $\min(\cdot,\cdot)$. These additional operators extends the hypothesis space by enabling piecewise-defined and bounded functional behavior. For example, the electronic band gap is non-negative by definition. Incorporating $\max$ and $\min$ directly into the symbolic form allows such bounds to emerge naturally within the learned expression. As shown in Ref.~\cite{Ma2025}, this can improve modeling of constrained quantities, including cumulative band-gap distributions.

Several prediction tasks involve probabilistic outputs, such as the likelihood that a material is metallic, insulating, or glass-forming. These quantities are restricted to the interval $[0,1]$. Rather than introducing task-specific output layers or activation functions, we represent these constraints within the same symbolic formalism. Regression and classification problems are therefore treated on equal footing. By extending the operator set to include bounded non-smooth functions, physically meaningful constraints become part of the learned analytical expression itself.

\subsection{Symbolic Regression}

\begin{figure}[thpb]
  \centering
  \includegraphics[width=.5\textwidth]{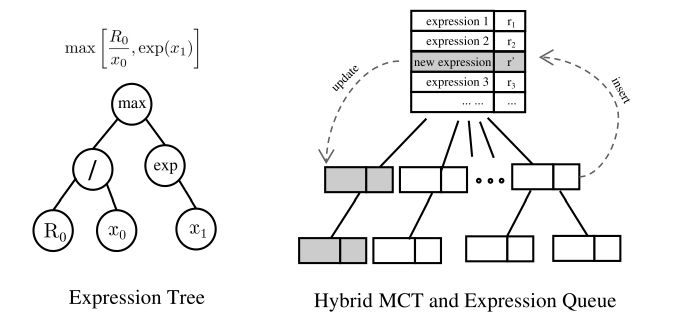}
  \caption{Hybrid Monte Carlo tree search and genetic programming.}
  \label{fig:hybrid}
\end{figure}

The proposed function class and expanded operator set introduce significant challenges for conventional Monte Carlo tree search (MCTS)-based symbolic regression.

First, the dimensionality of the continuous optimization problem increases substantially. In addition to the symbolic parameters $\bm{\theta}$, whose number is typically of order $\mathcal{O}(1)$, the model introduces elemental weights $\bm{w}$ whose dimensionality scales with the number of chemical species represented in the dataset. In practice, this corresponds to an additional $\mathcal{O}(10^2)$ trainable variables. The parameter-fitting stage therefore becomes a moderately high-dimensional nonlinear optimization problem, with a corresponding increase in computational cost.

Second, the inclusion of non-smooth operators such as $\max(\cdot,\cdot)$ and $\min(\cdot,\cdot)$ further complicates the search procedure. During the MCTS stage, enlarging the operator set increases the branching factor of the search tree, thereby expanding the combinatorial search space and increasing exploration cost. During parameter refinement, the resulting symbolic expressions are generally piecewise defined and non-differentiable at switching boundaries. The fitting problem is thus transformed into a segmented optimization landscape that can reduce convergence efficiency.

\subsubsection{Gradient-Based Optimization Strategy}

A gradient-based refinement strategy is adopted for continuous parameter optimization. Although operators such as $\max(\cdot,\cdot)$ and $\min(\cdot,\cdot)$ introduce non-smoothness, the resulting objective remains piecewise differentiable, with derivatives defined almost everywhere except at switching boundaries. This structure permits the practical use of gradient-based optimization methods.

For each candidate symbolic expression, the elemental weights $\bm{w}$ and symbolic coefficients $\bm{\theta}$ are optimized using the limited-memory Broyden--Fletcher--Goldfarb--Shanno (L-BFGS) algorithm~\cite{saputro2017}. To improve robustness with respect to initialization, we employ a multi-start strategy in which each expression is optimized from several independent initial conditions. In the calculations below, two random initializations and two positive random initializations are used, the latter ensuring validity in the presence of logarithmic terms. The solution with the lowest loss is retained.

This procedure is effective when candidate expressions exhibit moderate curvature and are locally close to convex within the physically relevant parameter region. More complex landscapes may still contain additional local minima, particularly for highly nonlinear expressions. Nevertheless, we assume that physically meaningful governing relations are typically compact and structured, allowing efficient convergence via local refinement.

\subsubsection{Hybrid MCTS and GP}

To improve search efficiency, we employ a hybrid Monte Carlo tree search--genetic programming (MCTS--GP) framework, extending recent symbolic-regression search strategies~\cite{huang2025,xu2023,landajuela2022,mundhenk2021}. The method combines the directed exploration of MCTS with the ``stage-jumping''\cite{huang2025} of GP, enabling efficient traversal of the enlarged symbolic and parametric search space.

In implementations of Ref.~\cite{huang2025}, candidate-expression queues are stored at many tree nodes. Here, we retain the global expression queue only at the root node, and all genetic operations---including mutation and crossover---are performed exclusively on this shared population. This substantially reduces memory overhead, since the number of tree nodes can grow rapidly during exploration. The root population continuously accumulates high-quality expressions discovered throughout the tree. Concentrating GP operations at the root therefore preserves diversity while allowing globally competitive candidates from different branches to recombine.

We further retain both backward and forward information propagation. During back-propagation, rewards obtained from expanded nodes are propagated to the root, and successful expressions are inserted into the root population. During forward propagation, newly accepted root expressions update the statistics of descendant nodes along their associated symbolic paths. Consequently, the MCTS tree and GP population evolve in a coordinated manner, with each component reinforcing the other.

This hybrid design improves sample efficiency, controls memory growth, and accelerates convergence toward analytical expressions. Full algorithmic details are provided in the Supporting Materials.

\subsubsection{Parallelism}
To further improve computational efficiency, we implement parallelism in both the GP and MCTS components of the framework.

In the GP stage, we select a batch of expressions from the root expression queue (with twice the target batch size to facilitate crossover operations). These expressions are then subjected to mutation or crossover to generate new candidate expressions in parallel. The subsequent parameter optimization for each candidate expression is performed in parallel, thereby reducing the time of the evolutionary refinement step.

In the MCTS stage, parallelism is introduced at the simulation phase. Specifically, we select and expand a batch of nodes simultaneously. For each expanded node, we perform rollout to generate candidate expressions and carry out parameter optimization in parallel. After evaluation, the corresponding rewards and expressions are back-propagated to update the search tree and the root expression queue.

\section{Results}
\subsection{Benchmarks}

We evaluate the proposed framework on three representative MatBench tasks~\cite{dunn2020}: \textit{matbench\_expt\_gap}, \textit{matbench\_expt\_is\_metal}, and \textit{matbench\_glass}, and compare against the MatBench v0.1 leaderboard.

\begin{table}[t]
  \centering
  \caption{Benchmark performance across three representative MatBench tasks\cite{dunn2020,matbench_expt_gap_leaderboard,matbench_expt_is_metal_leaderboard,matbench_glass_leaderboard}. Values in parentheses denote one standard deviation in the last digits. Lower MAE indicates better performance for band-gap prediction, whereas higher ROC-AUC indicates better performance for metallicity and glass classification. Model sizes are approximate parameter counts when available.}
  \label{tab:matbench}
  \small
  \begin{tabular}{lcccc}
    \hline
    \textbf{Model}
    & \begin{tabular}{c} Model \\ size \end{tabular}
    & \begin{tabular}{c} Band gap \\ MAE  \end{tabular}
    & \begin{tabular}{c} Metallicity \\ ROC-AUC \end{tabular}
    & \begin{tabular}{c} Glass \\ ROC-AUC  \end{tabular} \\
    \hline
    Darwin       & $\sim7$B      & \textbf{0.287(8)} & \textbf{0.960(4)} & 0.767(13) \\
    CrabNet      & $\sim10^6$      & 0.331(7)          & ---               & ---       \\
    MODNet       & $\sim10^6$    & 0.333(24)         & 0.916(7)          & \textbf{0.960(8)} \\
    AMMExpress   & N/R           & 0.416(19)         & 0.921(3)          & 0.861(20) \\
    RF-SCM       & N/R           & 0.446(18)         & 0.917(6)          & 0.859(16) \\
    GPTChem      & $\sim10^9$      & 0.454(12)         & 0.897(6)          & 0.776(12) \\
    Dummy        & 0             & 1.14(3)           & 0.492(13)         & 0.501(18) \\
    \hline
    ReLU model   & $\sim10^2$ & 0.575(36)    & ---               & ---       \\
    \textbf{Ours}& $\sim10^2$    & 0.471(23)         & 0.873(9)          & 0.816(14) \\
    \hline
  \end{tabular}
\end{table}

As summarized in Table~\ref{tab:matbench}, the proposed framework achieves competitive performance across all tasks while maintaining a fully explicit analytical form. The comparison includes non-analytical neural network-based models such as CrabNet~\cite{wang2021,wang2022} and MODNet~\cite{de2021a,de2021b}, large language model-based approaches such as Darwin~\cite{xie2023,xie2025} and GPTChem~\cite{jablonka2024}, conventional machine-learning methods such as RF-SCM/Magpie~\cite{ward2016,dunn2020}, and a simple analytical baseline proposed in Ref.~\cite{Ma2025},
\begin{equation}
  \label{eq:relu}
  P=\max\!\left(0,\sum_i w_i c_i\right),
\end{equation}
which corresponds to a minimal composition-weighted linear model with a non-negativity constraint.

Although the proposed method does not yet reach the accuracy of the strongest black-box models, the performance gap remains moderate given its fully symbolic and highly constrained functional form. Importantly, this is achieved with substantially fewer trainable parameters than typical neural-network architectures. LLM-scale models such as Darwin and GPTChem contain more than $10^9$ parameters~\cite{xie2023,xie2025,jablonka2024}. CrabNet is an attention-based composition model with approximately $10^6$ parameters~\cite{wang2021,wang2022}, while MODNet is a deep neural network with on the order of $10^6$ parameters~\cite{de2021a,de2021b}. In contrast, RF-SCM/Magpie is based on random forests over handcrafted descriptors and is effectively non-parametric in the neural-network sense~\cite{ward2016,dunn2020}. The ReLU baseline and the proposed method are fully analytical models, with parameters primarily consisting of elemental weights and a small number of fitted constants in the symbolic expression, resulting in an effective parameter count on the order of $10^2$.

\subsection{Discovered Expressions}

\begin{figure}[t]
  \centering
  \includegraphics[width=0.48\textwidth]{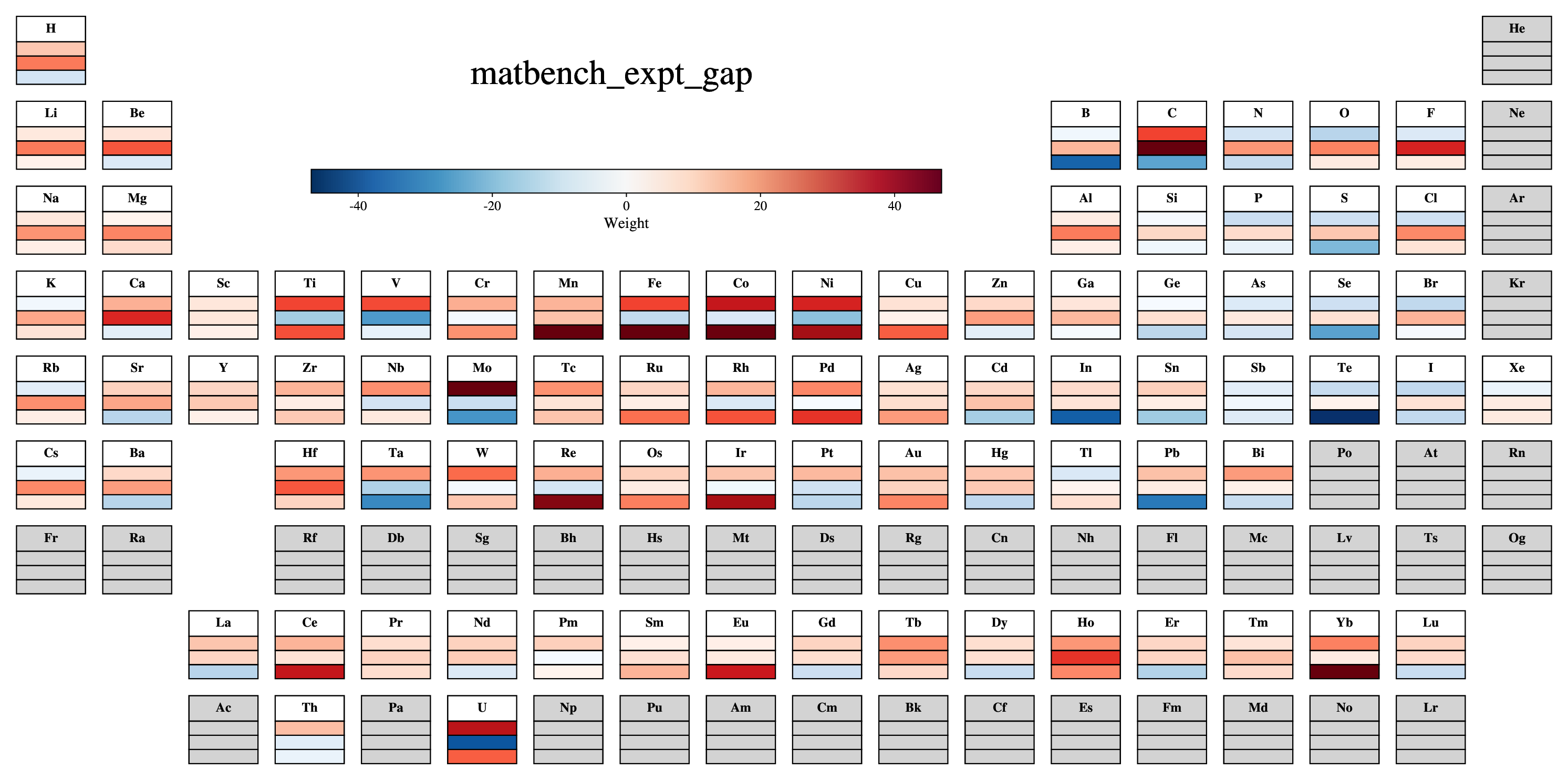}
  \includegraphics[width=0.48\textwidth]{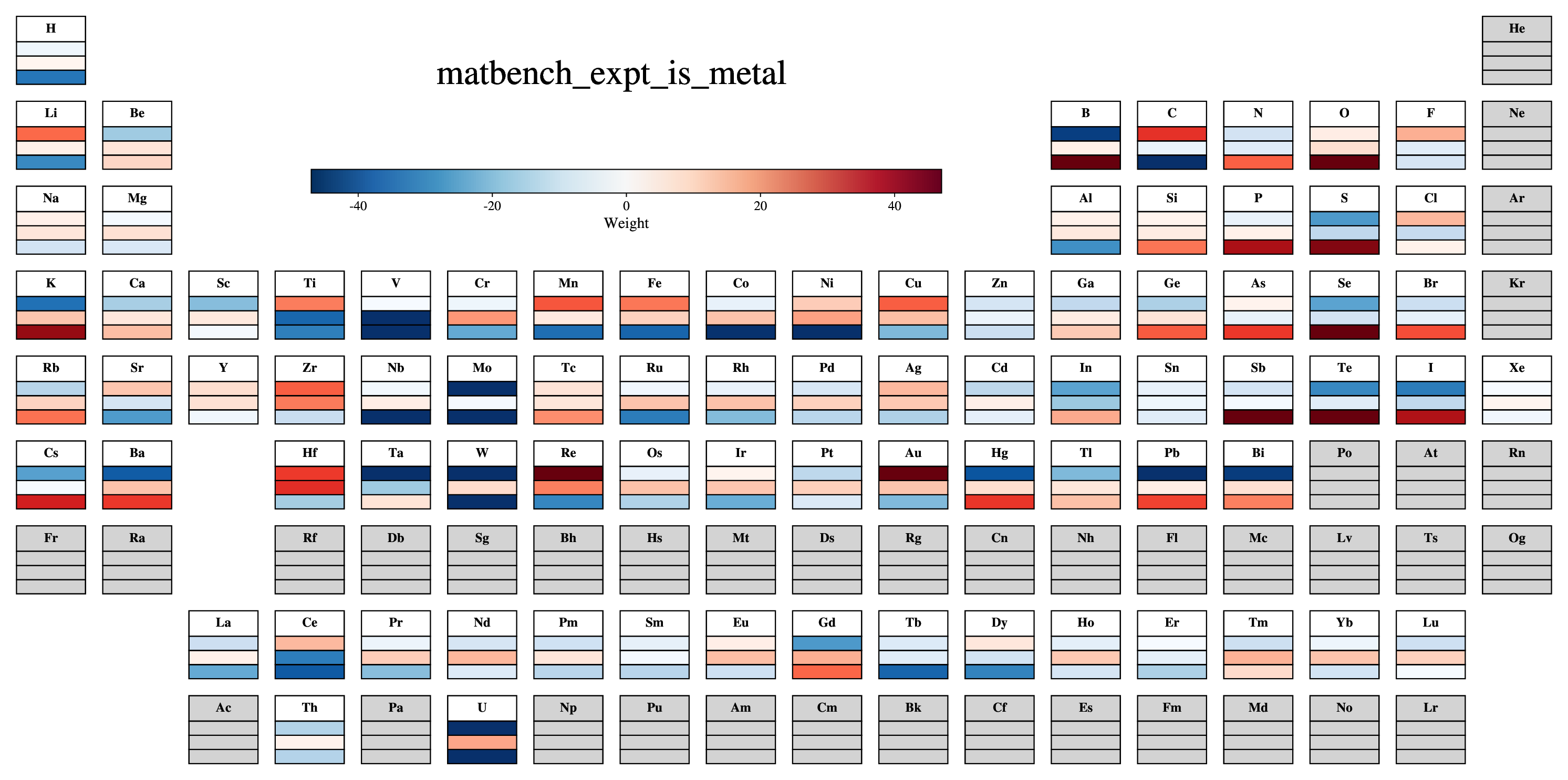}
  \includegraphics[width=0.48\textwidth]{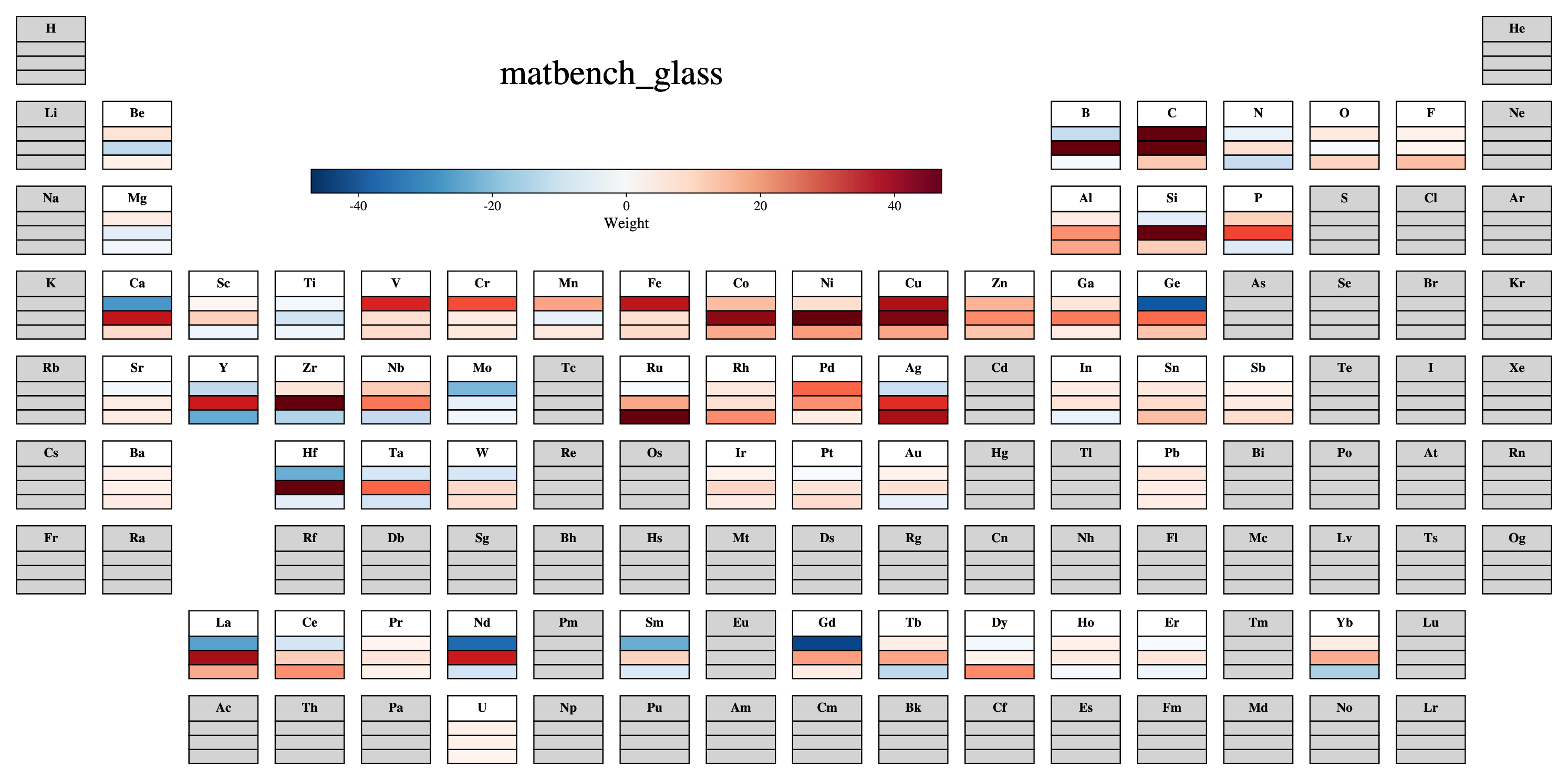}
  \caption{\label{fig:periodic}
    Learned elemental weights visualized on the periodic table for (top) \textit{matbench\_expt\_gap}, (middle) \textit{matbench\_expt\_is\_metal}, and (bottom) \textit{matbench\_glass}.}
\end{figure}

By fitting the full datasets, we obtain compact symbolic expressions for each target property, as shown in Eqs.~(\ref{eq:expt_gap})--(\ref{eq:glass}). Figure~\ref{fig:periodic} shows the corresponding learned elemental weights; the numerical values are provided in the Supporting Materials. The distinct periodic trends across tasks indicate that the model learns task-dependent effective elemental descriptors directly from data. Here $x_0$, $x_1$, and $x_2$ denote learned composition-weighted elemental variables defined in Eq.~\eqref{eq:general_form}. Although these expressions are not unique, they demonstrate that diverse material properties can be represented within a unified low-dimensional symbolic framework.

\begin{equation}
\label{eq:expt_gap}
\begin{aligned}
\mathcal{F}_{\mathrm{gap}} =\;&x_1 \exp\!\Big[-\exp\big(\max(x_2,\min(x_0,x_1))\big)\Big].
\end{aligned}
\end{equation}

\begin{equation}
\label{eq:expt_metal}
\begin{aligned}
\mathcal{F}_{\mathrm{metal}} =\;&\exp\Big[\min(-x_0,x_2)\,\exp(-\exp(x_1/2))\\
&\quad \times\exp(\min(x_0+x_2,x_1-x_2))\Big].
\end{aligned}
\end{equation}

\begin{equation}
\label{eq:glass}
\begin{aligned}
\mathcal{F}_{\mathrm{glass}} =\;&\exp\Big[(x_1-\exp(x_0\cdot\min(x_0,x_2)))\,\exp(-x_1)\Big].
\end{aligned}
\end{equation}

The discovered symbolic expressions share a common structural motif in which physical behavior emerges from competing elemental contributions mediated by extremal operators and nested nonlinearities.

The $\min$/$\max$ functions act as selector operators that identify limiting chemical environments, consistent with alloy systems in which the most restrictive elemental component can dominate the macroscopic response. Such behavior is physically plausible for composition-based properties, where a single unfavorable constituent may suppress conductivity, destabilize bonding networks, or constrain gap formation.

Nested exponential functions introduce hierarchical nonlinear gating. For example, terms of the form $\exp[-\exp(\cdot)]$ are bounded in the interval $(0,1]$ and strongly suppress the output once the internal descriptor exceeds a threshold-like value. This makes them naturally suitable for probability-related targets such as metallicity or glass formation. Interestingly, a similar structure also appears in the band-gap expression Eq.~\ref{eq:expt_gap}, suggesting that the model first performs an implicit classification between metallic and insulating systems through the nonlinear gating factor, while the prefactor $x_1$ primarily sets the magnitude of the finite gap once the system lies in the insulating regime.

\begin{figure}[t]
  \centering
  \includegraphics[width=.48\textwidth]{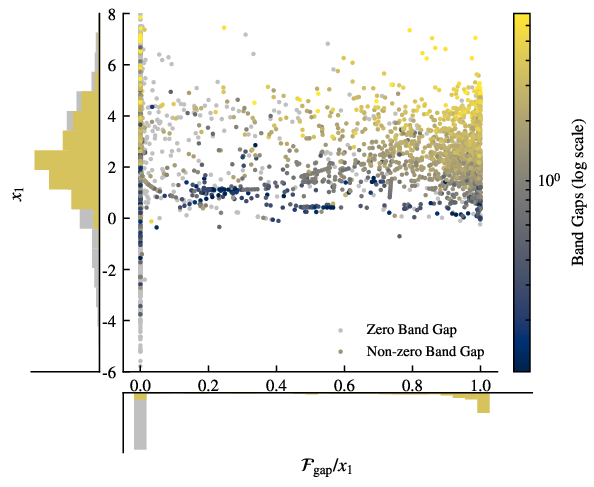}
  \caption{\label{fig:bg_exp}
    Scatter plot of $x_1$ versus $\mathcal{F}_{\rm gap}/x_1$ with aligned marginal histograms. Gray points denote zero-gap systems, while colored points represent finite-gap compounds (log scale). The distributions indicate separation between metallic and insulating materials.}
\end{figure}

\begin{figure*}[t]
  \centering
  \includegraphics[width=\textwidth]{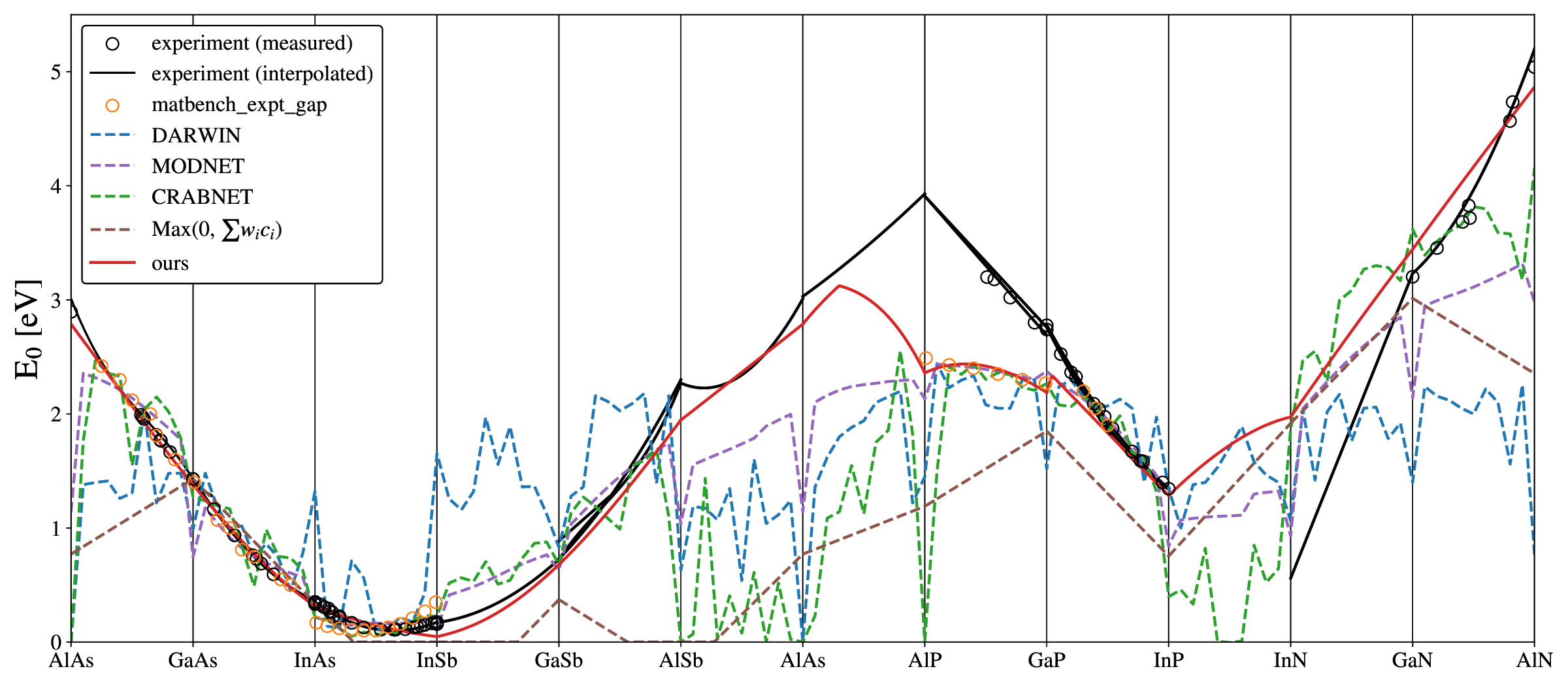}
\caption{\label{fig:ternary}
  Band gap comparison for selected III--V semiconductor alloys (adapted from Ref.~\cite{adachi2009}). Experimental data (open black circles) and interpolation fits (black solid line)~\cite{aubel1985,aspnes1986,saxena1981,monemar1976,kim2003,gaskill1990,woolley1964,dobbelaere1989,vishnubhatla1969,auvergne1974,roth1978,alibert1983,bignazzi1998,bellani1999,saadallah2003,rodriguez1991,choi2000,onton1971,alibert1972,schormann2006,mullhauser1998,goldhahn2000,takanobu2000,kasic2002}; MatBench training data~\cite{dunn2020} (open orange circles). Predictions: DARWIN~\cite{xie2023,xie2025} (cyan dashed), MODNet~\cite{de2021a,de2021b} (purple dashed), CrabNet~\cite{wang2021,wang2022} (green dashed), the ReLU model~\cite{Ma2025} (brown dashed) and this work (red solid). Data sources are detailed in the Supporting Materials.}
\end{figure*}

This interpretation is supported by Fig.~\ref{fig:bg_exp}, where $x_1$ is plotted against $\mathcal{F}_{\rm gap}/x_1$, corresponding to the isolated nonlinear gating contribution. Metallic systems with zero band gap cluster near strongly suppressed gating values, whereas insulating compounds occupy a distinct region with larger activation factors. The figure therefore suggests that the learned expression decomposes the problem into two coupled components: a latent metal--insulator discriminator and a continuous gap-scale predictor.

Across all tasks, the learned weights $w_0$, $w_1$, and $w_2$ act as latent elemental descriptors. The color gradients, ranging from negative (blue) to positive (red), reveal that elements with similar chemical and electronic characteristics tend to play analogous predictive roles for a given target property. Clear group-wise patterns emerge across the periodic table.

For example, in the \textit{matbench\_expt\_gap} task shown in Fig.~\ref{fig:periodic}, the halogens exhibit a characteristic sandwiched pattern across the three learned weights: the first and third are predominantly negative, while the second is positive. In contrast, several transition metals display an opposite trend. Elements such as Fe, Co, and Ni possess a negative middle weight flanked by two positive outer weights. These structured signatures indicate that chemically related elements are embedded into similar regions of the learned descriptor space.

Such behavior may reflect underlying periodic trends in real chemical properties. For halogens, the pattern is consistent with strong electronegativity and a tendency to stabilize ionic or insulating bonding environments, which often correlate with larger band gaps. For transition metals, the opposite tendency may be associated with partially filled $d$ orbitals, metallic bonding, and enhanced electronic delocalization, all of which generally favor smaller band gaps or metallic behavior.

Within each task, color similarity therefore reflects property-specific chemical similarity. However, the learned mappings are not universal across different targets. Elements such as oxygen display strongly positive weights for metallicity prediction, yet weaker or even negative contributions for band-gap and glass-forming tasks, indicating that elemental influence depends on both the target property and the surrounding symbolic functional form.

These results indicate that the symbolic-regression framework is able to recover interpretable periodic trends directly from data without predefined descriptors. The learned elemental weights provide a quantitative map of which regions of chemical space most strongly govern a given property, offering physically meaningful guidance for composition design and targeted experimental exploration.

\subsection{Band gap for selected III--V semiconductor alloys}

We further evaluate the model by predicting the band gap $E_0$ of selected III--V ternary semiconductor alloys, as shown in Fig.~\ref{fig:ternary}. The predictions are compared with experimental measurements and interpolation formulas, while the training data from the MatBench dataset are indicated for reference.

Systematic discrepancies are observed between the MatBench dataset and experimental values in certain systems, such as InAs--InSb and AlP--GaP. Since the model is trained on the MatBench dataset, it naturally reflects these inconsistencies, leading to deviations from experimentally measured band gaps. For example, in the AlP--GaP system, the dataset reports significantly smaller band gaps than experiment. Consequently, the model reproduces this trend and underestimates the band gaps of Al-based compounds (e.g., AlP-AlAs and AlP-GaP).

Despite these limitations, the model captures the overall band-gap landscape across the full composition range. It correctly reproduces the global trends, including the decrease from AlAs to InSb, the increase toward AlP. In regions with training data, the predictions closely match the dataset, while in data-sparse regions the model provides smooth and physically reasonable interpolation.

We further compare with models, including Darwin, CrabNet, MODNet and the ReLU model Eq.~\ref{eq:relu}. While these neural networ based models achieve lower MAE on benchmark datasets, their predictions exhibit notable limitations in compositional interpolation. First, they perform well at compositions present in the training set but can show inconsistent predictions for nearby unseen compositions. Second, in regions without training data, their predictions are less reliable. Third, and most importantly, these models often produce discontinuous or fluctuating band-gap profiles as a function of composition. For example, abrupt variations are observed between neighboring compositions in Darwin and CrabNet, while MODNet, although smoother, can still introduce discontinuities due to feature changes (e.g., the transition from binary to ternary compositions).

In contrast, the proposed symbolic regression model produces a smooth and continuous band-gap profile across the entire compositional space. This behavior arises naturally from the analytical functional form and provides a physically consistent description of composition-dependent properties. The resulting profiles exhibit continuous behavior with respect to composition, in contrast to the discontinuities observed in competing models.

\subsection{Limitations}

Despite its advantages, the proposed approach has several limitations. First, although symbolic regression yields explicit analytical expressions, interpretability is not guaranteed. The discovered expressions can become highly complex, requiring additional analysis to extract meaningful physical insights and potentially undermining their practical interpretability.

Second, the optimization procedure is inherently approximate. Monte Carlo tree search (MCTS) provides high-quality but generally suboptimal solutions and does not guarantee global optimality. For discounted Markov decision processes (MDPs) with discount factor $0<\gamma<1$, obtaining an $\varepsilon$-optimal action requires a search tree whose depth scales as $\mathcal{O}\!\left(\frac{1}{1-\gamma}\log\frac{1}{\varepsilon(1-\gamma)}\right)$ and width as $\mathcal{O}\!\left(\frac{K}{\varepsilon(1-\gamma)}\right)$, where $K$ denotes the number of actions~\cite{kearns2002,kocsis2006}. In addition, the gradient-based optimization stage is inherently local; despite multi-start initialization strategies, convergence to the global optimum is not guaranteed. In Section S3 of the Supporting Information, we list the top five candidate expressions discovered by the search procedure, and in Section S6 we evaluate their behavior for III--V alloy band-gap prediction.

Third, the method is susceptible to overfitting in low-data regimes. In datasets with limited samples, such as the MatBench steel dataset (312 samples), the flexibility of symbolic regression can lead to overly complex expressions rather than underlying physical trends.

\section{Discussion}

We have introduced a composition-weighted symbolic regression framework for general-purpose composition-based materials property prediction. The method combines interpretable modeling with data-driven function discovery by learning both analytical functional forms and elemental weightings directly from data, without relying on predefined descriptors. Another advantage of the approach is its natural incorporation of physical constraints through operators such as $\max$ and $\min$. These enable bounded or piecewise behavior to be encoded directly in the symbolic expression, allowing non-negative quantities (e.g., band gaps) and probabilistic outputs in $[0,1]$ to be modeled within a unified formalism, without task-specific output layers.

Benchmark evaluations on representative MatBench tasks demonstrate that the proposed model achieves comparable accuracy relative to state-of-the-art black-box methods, despite using substantially fewer trainable parameters. In addition, the closed-form expressions provide smooth and physically consistent predictions across continuous composition spaces, which is valuable for interpolation and extrapolation in data-sparse regimes, as illustrated for III--V semiconductor alloys. The learned elemental weights also exhibit chemically meaningful periodic trends, suggesting that the model can recover task-relevant latent descriptors directly from data.

Several limitations remain. First, symbolic expressions are explicit but not always simple; highly accurate solutions may still be algebraically complex and require further interpretation. Second, the hybrid MCTS--GP optimization strategy is approximate and does not guarantee global optimality, while computational cost increases with search depth and elemental diversity. Third, the flexibility of symbolic regression can lead to overfitting in low-data regimes. Finally, the present functional space may be insufficient for strongly multiscale or sharply discontinuous phenomena, such as complex phase-boundary behavior in metallic-glass systems.

\section{Data availability}
The code, scripts, and data supporting the findings of this study are available at \url{https://github.com/yangh618/Composition-weighted-SR_matbench.git}. A general implementation of the composition-weighted symbolic regression framework is available at \url{https://github.com/yangh618/Composition-weighted-SR.git}.

\section*{Acknowledgements}
This work was partially supported by NSFC grant 12425113.

\bibliography{refs}     

\begin{thebibliography}{59}%
\makeatletter
\providecommand \@ifxundefined [1]{%
 \@ifx{#1\undefined}
}%
\providecommand \@ifnum [1]{%
 \ifnum #1\expandafter \@firstoftwo
 \else \expandafter \@secondoftwo
 \fi
}%
\providecommand \@ifx [1]{%
 \ifx #1\expandafter \@firstoftwo
 \else \expandafter \@secondoftwo
 \fi
}%
\providecommand \natexlab [1]{#1}%
\providecommand \enquote  [1]{``#1''}%
\providecommand \bibnamefont  [1]{#1}%
\providecommand \bibfnamefont [1]{#1}%
\providecommand \citenamefont [1]{#1}%
\providecommand \href@noop [0]{\@secondoftwo}%
\providecommand \href [0]{\begingroup \@sanitize@url \@href}%
\providecommand \@href[1]{\@@startlink{#1}\@@href}%
\providecommand \@@href[1]{\endgroup#1\@@endlink}%
\providecommand \@sanitize@url [0]{\catcode `\\12\catcode `\$12\catcode
  `\&12\catcode `\#12\catcode `\^12\catcode `\_12\catcode `\%12\relax}%
\providecommand \@@startlink[1]{}%
\providecommand \@@endlink[0]{}%
\providecommand \url  [0]{\begingroup\@sanitize@url \@url }%
\providecommand \@url [1]{\endgroup\@href {#1}{\urlprefix }}%
\providecommand \urlprefix  [0]{URL }%
\providecommand \Eprint [0]{\href }%
\providecommand \doibase [0]{https://doi.org/}%
\providecommand \selectlanguage [0]{\@gobble}%
\providecommand \bibinfo  [0]{\@secondoftwo}%
\providecommand \bibfield  [0]{\@secondoftwo}%
\providecommand \translation [1]{[#1]}%
\providecommand \BibitemOpen [0]{}%
\providecommand \bibitemStop [0]{}%
\providecommand \bibitemNoStop [0]{.\EOS\space}%
\providecommand \EOS [0]{\spacefactor3000\relax}%
\providecommand \BibitemShut  [1]{\csname bibitem#1\endcsname}%
\let\auto@bib@innerbib\@empty
\bibitem [{\citenamefont {Xie}\ and\ \citenamefont {Grossman}(2018)}]{xie2018}%
  \BibitemOpen
  \bibfield  {author} {\bibinfo {author} {\bibfnamefont {T.}~\bibnamefont
  {Xie}}\ and\ \bibinfo {author} {\bibfnamefont {J.~C.}\ \bibnamefont
  {Grossman}},\ }\bibfield  {title} {\bibinfo {title} {{C}rystal graph
  convolutional neural networks for an accurate and interpretable prediction of
  material properties},\ }\href@noop {} {\bibfield  {journal} {\bibinfo
  {journal} {Phys. Rev. Lett.}\ }\textbf {\bibinfo {volume} {120}},\ \bibinfo
  {pages} {145301} (\bibinfo {year} {2018})}\BibitemShut {NoStop}%
\bibitem [{\citenamefont {Sch{\"u}tt}\ \emph {et~al.}(2018)\citenamefont
  {Sch{\"u}tt}, \citenamefont {Sauceda}, \citenamefont {Kindermans},
  \citenamefont {Tkatchenko},\ and\ \citenamefont {M{\"u}ller}}]{schutt2018}%
  \BibitemOpen
  \bibfield  {author} {\bibinfo {author} {\bibfnamefont {K.~T.}\ \bibnamefont
  {Sch{\"u}tt}}, \bibinfo {author} {\bibfnamefont {H.~E.}\ \bibnamefont
  {Sauceda}}, \bibinfo {author} {\bibfnamefont {P.-J.}\ \bibnamefont
  {Kindermans}}, \bibinfo {author} {\bibfnamefont {A.}~\bibnamefont
  {Tkatchenko}},\ and\ \bibinfo {author} {\bibfnamefont {K.-R.}\ \bibnamefont
  {M{\"u}ller}},\ }\bibfield  {title} {\bibinfo {title} {{S}chnet--a deep
  learning architecture for molecules and materials},\ }\href@noop {}
  {\bibfield  {journal} {\bibinfo  {journal} {J. Chem. Phys.}\ }\textbf
  {\bibinfo {volume} {148}} (\bibinfo {year} {2018})}\BibitemShut {NoStop}%
\bibitem [{\citenamefont {Chen}\ \emph {et~al.}(2019)\citenamefont {Chen},
  \citenamefont {Ye}, \citenamefont {Zuo}, \citenamefont {Zheng},\ and\
  \citenamefont {Ong}}]{chen2019}%
  \BibitemOpen
  \bibfield  {author} {\bibinfo {author} {\bibfnamefont {C.}~\bibnamefont
  {Chen}}, \bibinfo {author} {\bibfnamefont {W.}~\bibnamefont {Ye}}, \bibinfo
  {author} {\bibfnamefont {Y.}~\bibnamefont {Zuo}}, \bibinfo {author}
  {\bibfnamefont {C.}~\bibnamefont {Zheng}},\ and\ \bibinfo {author}
  {\bibfnamefont {S.~P.}\ \bibnamefont {Ong}},\ }\bibfield  {title} {\bibinfo
  {title} {{G}raph networks as a universal machine learning framework for
  molecules and crystals},\ }\href@noop {} {\bibfield  {journal} {\bibinfo
  {journal} {Chem. Mater.}\ }\textbf {\bibinfo {volume} {31}},\ \bibinfo
  {pages} {3564} (\bibinfo {year} {2019})}\BibitemShut {NoStop}%
\bibitem [{\citenamefont {Gasteiger}\ \emph {et~al.}(2020)\citenamefont
  {Gasteiger}, \citenamefont {Gro{\ss}},\ and\ \citenamefont
  {G{\"u}nnemann}}]{gasteiger2020}%
  \BibitemOpen
  \bibfield  {author} {\bibinfo {author} {\bibfnamefont {J.}~\bibnamefont
  {Gasteiger}}, \bibinfo {author} {\bibfnamefont {J.}~\bibnamefont
  {Gro{\ss}}},\ and\ \bibinfo {author} {\bibfnamefont {S.}~\bibnamefont
  {G{\"u}nnemann}},\ }\bibfield  {title} {\bibinfo {title} {{D}irectional
  message passing for molecular graphs},\ }\href@noop {} {\bibfield  {journal}
  {\bibinfo  {journal} {arXiv preprint arXiv:2003.03123}\ } (\bibinfo {year}
  {2020})}\BibitemShut {NoStop}%
\bibitem [{\citenamefont {Dunn}\ \emph {et~al.}(2020)\citenamefont {Dunn},
  \citenamefont {Wang}, \citenamefont {Ganose}, \citenamefont {Dopp},\ and\
  \citenamefont {Jain}}]{dunn2020}%
  \BibitemOpen
  \bibfield  {author} {\bibinfo {author} {\bibfnamefont {A.}~\bibnamefont
  {Dunn}}, \bibinfo {author} {\bibfnamefont {Q.}~\bibnamefont {Wang}}, \bibinfo
  {author} {\bibfnamefont {A.}~\bibnamefont {Ganose}}, \bibinfo {author}
  {\bibfnamefont {D.}~\bibnamefont {Dopp}},\ and\ \bibinfo {author}
  {\bibfnamefont {A.}~\bibnamefont {Jain}},\ }\bibfield  {title} {\bibinfo
  {title} {{B}enchmarking materials property prediction methods: the {M}atbench
  test set and {A}utomatminer reference algorithm},\ }\href@noop {} {\bibfield
  {journal} {\bibinfo  {journal} {Npj Comput. Mater.}\ }\textbf {\bibinfo
  {volume} {6}},\ \bibinfo {pages} {138} (\bibinfo {year} {2020})}\BibitemShut
  {NoStop}%
\bibitem [{\citenamefont {Liao}\ and\ \citenamefont {Smidt}(2022)}]{liao2022}%
  \BibitemOpen
  \bibfield  {author} {\bibinfo {author} {\bibfnamefont {Y.-L.}\ \bibnamefont
  {Liao}}\ and\ \bibinfo {author} {\bibfnamefont {T.}~\bibnamefont {Smidt}},\
  }\bibfield  {title} {\bibinfo {title} {{E}quiformer: {E}quivariant graph
  attention transformer for 3d atomistic graphs},\ }\href@noop {} {\bibfield
  {journal} {\bibinfo  {journal} {arXiv preprint arXiv:2206.11990}\ } (\bibinfo
  {year} {2022})}\BibitemShut {NoStop}%
\bibitem [{\citenamefont {Kim}\ \emph {et~al.}(2026)\citenamefont {Kim},
  \citenamefont {You}, \citenamefont {Park}, \citenamefont {Lim}, \citenamefont
  {Kang}, \citenamefont {Kim}, \citenamefont {Jeon}, \citenamefont {Ju},
  \citenamefont {Hong}, \citenamefont {Lee} \emph {et~al.}}]{kim2026}%
  \BibitemOpen
  \bibfield  {author} {\bibinfo {author} {\bibfnamefont {J.}~\bibnamefont
  {Kim}}, \bibinfo {author} {\bibfnamefont {J.}~\bibnamefont {You}}, \bibinfo
  {author} {\bibfnamefont {Y.}~\bibnamefont {Park}}, \bibinfo {author}
  {\bibfnamefont {Y.}~\bibnamefont {Lim}}, \bibinfo {author} {\bibfnamefont
  {Y.}~\bibnamefont {Kang}}, \bibinfo {author} {\bibfnamefont {J.}~\bibnamefont
  {Kim}}, \bibinfo {author} {\bibfnamefont {H.}~\bibnamefont {Jeon}}, \bibinfo
  {author} {\bibfnamefont {S.}~\bibnamefont {Ju}}, \bibinfo {author}
  {\bibfnamefont {D.}~\bibnamefont {Hong}}, \bibinfo {author} {\bibfnamefont
  {S.~Y.}\ \bibnamefont {Lee}}, \emph {et~al.},\ }\bibfield  {title} {\bibinfo
  {title} {{O}ptimizing cross-domain transfer for universal machine learning
  interatomic potentials},\ }\href@noop {} {\bibfield  {journal} {\bibinfo
  {journal} {Nat. Commun.}\ } (\bibinfo {year} {2026})}\BibitemShut {NoStop}%
\bibitem [{\citenamefont {Park}\ \emph {et~al.}(2024)\citenamefont {Park},
  \citenamefont {Kim}, \citenamefont {Hwang},\ and\ \citenamefont
  {Han}}]{park2024}%
  \BibitemOpen
  \bibfield  {author} {\bibinfo {author} {\bibfnamefont {Y.}~\bibnamefont
  {Park}}, \bibinfo {author} {\bibfnamefont {J.}~\bibnamefont {Kim}}, \bibinfo
  {author} {\bibfnamefont {S.}~\bibnamefont {Hwang}},\ and\ \bibinfo {author}
  {\bibfnamefont {S.}~\bibnamefont {Han}},\ }\bibfield  {title} {\bibinfo
  {title} {{S}calable parallel algorithm for graph neural network interatomic
  potentials in molecular dynamics simulations},\ }\href@noop {} {\bibfield
  {journal} {\bibinfo  {journal} {J. Chem. Theory Comput.}\ }\textbf {\bibinfo
  {volume} {20}},\ \bibinfo {pages} {4857} (\bibinfo {year}
  {2024})}\BibitemShut {NoStop}%
\bibitem [{\citenamefont {Zhang}\ \emph
  {et~al.}(2024{\natexlab{a}})\citenamefont {Zhang}, \citenamefont {Liu},
  \citenamefont {Zhang}, \citenamefont {Zhang}, \citenamefont {Cai},
  \citenamefont {Bi}, \citenamefont {Du}, \citenamefont {Qin}, \citenamefont
  {Peng}, \citenamefont {Huang} \emph {et~al.}}]{zhang2024}%
  \BibitemOpen
  \bibfield  {author} {\bibinfo {author} {\bibfnamefont {D.}~\bibnamefont
  {Zhang}}, \bibinfo {author} {\bibfnamefont {X.}~\bibnamefont {Liu}}, \bibinfo
  {author} {\bibfnamefont {X.}~\bibnamefont {Zhang}}, \bibinfo {author}
  {\bibfnamefont {C.}~\bibnamefont {Zhang}}, \bibinfo {author} {\bibfnamefont
  {C.}~\bibnamefont {Cai}}, \bibinfo {author} {\bibfnamefont {H.}~\bibnamefont
  {Bi}}, \bibinfo {author} {\bibfnamefont {Y.}~\bibnamefont {Du}}, \bibinfo
  {author} {\bibfnamefont {X.}~\bibnamefont {Qin}}, \bibinfo {author}
  {\bibfnamefont {A.}~\bibnamefont {Peng}}, \bibinfo {author} {\bibfnamefont
  {J.}~\bibnamefont {Huang}}, \emph {et~al.},\ }\bibfield  {title} {\bibinfo
  {title} {{D}{P}{A}-2: a large atomic model as a multi-task learner},\
  }\href@noop {} {\bibfield  {journal} {\bibinfo  {journal} {Npj Comput.
  Mater.}\ }\textbf {\bibinfo {volume} {10}},\ \bibinfo {pages} {293} (\bibinfo
  {year} {2024}{\natexlab{a}})}\BibitemShut {NoStop}%
\bibitem [{\citenamefont {Zhang}\ \emph
  {et~al.}(2024{\natexlab{b}})\citenamefont {Zhang}, \citenamefont {Bi},
  \citenamefont {Dai}, \citenamefont {Jiang}, \citenamefont {Liu},
  \citenamefont {Zhang},\ and\ \citenamefont {Wang}}]{zhang2024pretraining}%
  \BibitemOpen
  \bibfield  {author} {\bibinfo {author} {\bibfnamefont {D.}~\bibnamefont
  {Zhang}}, \bibinfo {author} {\bibfnamefont {H.}~\bibnamefont {Bi}}, \bibinfo
  {author} {\bibfnamefont {F.-Z.}\ \bibnamefont {Dai}}, \bibinfo {author}
  {\bibfnamefont {W.}~\bibnamefont {Jiang}}, \bibinfo {author} {\bibfnamefont
  {X.}~\bibnamefont {Liu}}, \bibinfo {author} {\bibfnamefont {L.}~\bibnamefont
  {Zhang}},\ and\ \bibinfo {author} {\bibfnamefont {H.}~\bibnamefont {Wang}},\
  }\bibfield  {title} {\bibinfo {title} {{P}retraining of attention-based deep
  learning potential model for molecular simulation},\ }\href@noop {}
  {\bibfield  {journal} {\bibinfo  {journal} {Npj Comput. Mater.}\ }\textbf
  {\bibinfo {volume} {10}},\ \bibinfo {pages} {94} (\bibinfo {year}
  {2024}{\natexlab{b}})}\BibitemShut {NoStop}%
\bibitem [{\citenamefont {De~Breuck}\ \emph
  {et~al.}(2021{\natexlab{a}})\citenamefont {De~Breuck}, \citenamefont
  {Hautier},\ and\ \citenamefont {Rignanese}}]{de2021a}%
  \BibitemOpen
  \bibfield  {author} {\bibinfo {author} {\bibfnamefont {P.-P.}\ \bibnamefont
  {De~Breuck}}, \bibinfo {author} {\bibfnamefont {G.}~\bibnamefont {Hautier}},\
  and\ \bibinfo {author} {\bibfnamefont {G.-M.}\ \bibnamefont {Rignanese}},\
  }\bibfield  {title} {\bibinfo {title} {{M}aterials property prediction for
  limited datasets enabled by feature selection and joint learning with
  {M}{O}{D}{N}et},\ }\href@noop {} {\bibfield  {journal} {\bibinfo  {journal}
  {Npj Comput. Mater.}\ }\textbf {\bibinfo {volume} {7}},\ \bibinfo {pages}
  {83} (\bibinfo {year} {2021}{\natexlab{a}})}\BibitemShut {NoStop}%
\bibitem [{\citenamefont {De~Breuck}\ \emph
  {et~al.}(2021{\natexlab{b}})\citenamefont {De~Breuck}, \citenamefont
  {Evans},\ and\ \citenamefont {Rignanese}}]{de2021b}%
  \BibitemOpen
  \bibfield  {author} {\bibinfo {author} {\bibfnamefont {P.-P.}\ \bibnamefont
  {De~Breuck}}, \bibinfo {author} {\bibfnamefont {M.~L.}\ \bibnamefont
  {Evans}},\ and\ \bibinfo {author} {\bibfnamefont {G.-M.}\ \bibnamefont
  {Rignanese}},\ }\bibfield  {title} {\bibinfo {title} {{R}obust model
  benchmarking and bias-imbalance in data-driven materials science: a case
  study on {M}{O}{D}{N}et},\ }\href@noop {} {\bibfield  {journal} {\bibinfo
  {journal} {J. Phys. Condens. Matter}\ }\textbf {\bibinfo {volume} {33}},\
  \bibinfo {pages} {404002} (\bibinfo {year} {2021}{\natexlab{b}})}\BibitemShut
  {NoStop}%
\bibitem [{\citenamefont {Ruff}\ \emph {et~al.}(2024)\citenamefont {Ruff},
  \citenamefont {Reiser}, \citenamefont {St{\"u}hmer},\ and\ \citenamefont
  {Friederich}}]{ruff2024}%
  \BibitemOpen
  \bibfield  {author} {\bibinfo {author} {\bibfnamefont {R.}~\bibnamefont
  {Ruff}}, \bibinfo {author} {\bibfnamefont {P.}~\bibnamefont {Reiser}},
  \bibinfo {author} {\bibfnamefont {J.}~\bibnamefont {St{\"u}hmer}},\ and\
  \bibinfo {author} {\bibfnamefont {P.}~\bibnamefont {Friederich}},\ }\bibfield
   {title} {\bibinfo {title} {{C}onnectivity optimized nested line graph
  networks for crystal structures},\ }\href@noop {} {\bibfield  {journal}
  {\bibinfo  {journal} {Digit. Discov.}\ }\textbf {\bibinfo {volume} {3}},\
  \bibinfo {pages} {594} (\bibinfo {year} {2024})}\BibitemShut {NoStop}%
\bibitem [{\citenamefont {Ihalage}\ and\ \citenamefont
  {Hao}(2022)}]{ihalage2022}%
  \BibitemOpen
  \bibfield  {author} {\bibinfo {author} {\bibfnamefont {A.}~\bibnamefont
  {Ihalage}}\ and\ \bibinfo {author} {\bibfnamefont {Y.}~\bibnamefont {Hao}},\
  }\bibfield  {title} {\bibinfo {title} {{F}ormula {G}raph {S}elf-{A}ttention
  {N}etwork for {R}epresentation-{D}omain {I}ndependent {M}aterials
  {D}iscovery},\ }\href@noop {} {\bibfield  {journal} {\bibinfo  {journal}
  {Adv. Sci.}\ }\textbf {\bibinfo {volume} {9}},\ \bibinfo {pages} {2200164}
  (\bibinfo {year} {2022})}\BibitemShut {NoStop}%
\bibitem [{\citenamefont {Wang}\ \emph {et~al.}(2021)\citenamefont {Wang},
  \citenamefont {Kauwe}, \citenamefont {Murdock},\ and\ \citenamefont
  {Sparks}}]{wang2021}%
  \BibitemOpen
  \bibfield  {author} {\bibinfo {author} {\bibfnamefont {A.~Y.-T.}\
  \bibnamefont {Wang}}, \bibinfo {author} {\bibfnamefont {S.~K.}\ \bibnamefont
  {Kauwe}}, \bibinfo {author} {\bibfnamefont {R.~J.}\ \bibnamefont {Murdock}},\
  and\ \bibinfo {author} {\bibfnamefont {T.~D.}\ \bibnamefont {Sparks}},\
  }\bibfield  {title} {\bibinfo {title} {{C}ompositionally restricted
  attention-based network for materials property predictions},\ }\href
  {https://doi.org/10.1038/s41524-021-00545-1} {\bibfield  {journal} {\bibinfo
  {journal} {Npj Comput. Mater.}\ }\textbf {\bibinfo {volume} {7}},\ \bibinfo
  {pages} {77} (\bibinfo {year} {2021})}\BibitemShut {NoStop}%
\bibitem [{\citenamefont {Wang}\ \emph {et~al.}(2022)\citenamefont {Wang},
  \citenamefont {Mahmoud}, \citenamefont {Czasny},\ and\ \citenamefont
  {Gurlo}}]{wang2022}%
  \BibitemOpen
  \bibfield  {author} {\bibinfo {author} {\bibfnamefont {A.~Y.-T.}\
  \bibnamefont {Wang}}, \bibinfo {author} {\bibfnamefont {M.~S.}\ \bibnamefont
  {Mahmoud}}, \bibinfo {author} {\bibfnamefont {M.}~\bibnamefont {Czasny}},\
  and\ \bibinfo {author} {\bibfnamefont {A.}~\bibnamefont {Gurlo}},\ }\bibfield
   {title} {\bibinfo {title} {{C}rab{N}et for {E}xplainable {D}eep {L}earning
  in {M}aterials {S}cience: {B}ridging the {G}ap {B}etween {A}cademia and
  {I}ndustry},\ }\href {https://doi.org/10.1007/s40192-021-00247-y} {\bibfield
  {journal} {\bibinfo  {journal} {Integr. Mater. Manuf. Innov.}\ }\textbf
  {\bibinfo {volume} {11}},\ \bibinfo {pages} {41} (\bibinfo {year}
  {2022})}\BibitemShut {NoStop}%
\bibitem [{\citenamefont {Meredig}\ \emph {et~al.}(2014)\citenamefont
  {Meredig}, \citenamefont {Agrawal}, \citenamefont {Kirklin}, \citenamefont
  {Saal}, \citenamefont {Doak}, \citenamefont {Thompson}, \citenamefont
  {Zhang}, \citenamefont {Choudhary},\ and\ \citenamefont
  {Wolverton}}]{meredig2014}%
  \BibitemOpen
  \bibfield  {author} {\bibinfo {author} {\bibfnamefont {B.}~\bibnamefont
  {Meredig}}, \bibinfo {author} {\bibfnamefont {A.}~\bibnamefont {Agrawal}},
  \bibinfo {author} {\bibfnamefont {S.}~\bibnamefont {Kirklin}}, \bibinfo
  {author} {\bibfnamefont {J.~E.}\ \bibnamefont {Saal}}, \bibinfo {author}
  {\bibfnamefont {J.~W.}\ \bibnamefont {Doak}}, \bibinfo {author}
  {\bibfnamefont {A.}~\bibnamefont {Thompson}}, \bibinfo {author}
  {\bibfnamefont {K.}~\bibnamefont {Zhang}}, \bibinfo {author} {\bibfnamefont
  {A.}~\bibnamefont {Choudhary}},\ and\ \bibinfo {author} {\bibfnamefont
  {C.}~\bibnamefont {Wolverton}},\ }\bibfield  {title} {\bibinfo {title}
  {{C}ombinatorial screening for new materials in unconstrained composition
  space with machine learning},\ }\href@noop {} {\bibfield  {journal} {\bibinfo
   {journal} {Phys. Rev. B}\ }\textbf {\bibinfo {volume} {89}},\ \bibinfo
  {pages} {094104} (\bibinfo {year} {2014})}\BibitemShut {NoStop}%
\bibitem [{\citenamefont {Ward}\ \emph {et~al.}(2016)\citenamefont {Ward},
  \citenamefont {Agrawal}, \citenamefont {Choudhary},\ and\ \citenamefont
  {Wolverton}}]{ward2016}%
  \BibitemOpen
  \bibfield  {author} {\bibinfo {author} {\bibfnamefont {L.}~\bibnamefont
  {Ward}}, \bibinfo {author} {\bibfnamefont {A.}~\bibnamefont {Agrawal}},
  \bibinfo {author} {\bibfnamefont {A.}~\bibnamefont {Choudhary}},\ and\
  \bibinfo {author} {\bibfnamefont {C.}~\bibnamefont {Wolverton}},\ }\bibfield
  {title} {\bibinfo {title} {{A} general-purpose machine learning framework for
  predicting properties of inorganic materials},\ }\href@noop {} {\bibfield
  {journal} {\bibinfo  {journal} {Npj Comput. Mater.}\ }\textbf {\bibinfo
  {volume} {2}},\ \bibinfo {pages} {16028} (\bibinfo {year}
  {2016})}\BibitemShut {NoStop}%
\bibitem [{\citenamefont {Goodall}\ and\ \citenamefont
  {Lee}(2020)}]{goodall2020}%
  \BibitemOpen
  \bibfield  {author} {\bibinfo {author} {\bibfnamefont {R.~E.~A.}\
  \bibnamefont {Goodall}}\ and\ \bibinfo {author} {\bibfnamefont {A.~A.}\
  \bibnamefont {Lee}},\ }\bibfield  {title} {\bibinfo {title} {{P}redicting
  {M}aterials {P}roperties without {C}rystal {S}tructure: {D}eep
  {R}epresentation {L}earning from {S}toichiometry},\ }\href@noop {} {\bibfield
   {journal} {\bibinfo  {journal} {Nat. Commun.}\ }\textbf {\bibinfo {volume}
  {11}},\ \bibinfo {pages} {6280} (\bibinfo {year} {2020})}\BibitemShut
  {NoStop}%
\bibitem [{\citenamefont {Guo}\ \emph {et~al.}(2022)\citenamefont {Guo},
  \citenamefont {Hu}, \citenamefont {Han},\ and\ \citenamefont
  {Ouyang}}]{guo2022}%
  \BibitemOpen
  \bibfield  {author} {\bibinfo {author} {\bibfnamefont {Z.}~\bibnamefont
  {Guo}}, \bibinfo {author} {\bibfnamefont {S.}~\bibnamefont {Hu}}, \bibinfo
  {author} {\bibfnamefont {Z.-K.}\ \bibnamefont {Han}},\ and\ \bibinfo {author}
  {\bibfnamefont {R.}~\bibnamefont {Ouyang}},\ }\bibfield  {title} {\bibinfo
  {title} {{I}mproving symbolic regression for predicting materials properties
  with iterative variable selection},\ }\href@noop {} {\bibfield  {journal}
  {\bibinfo  {journal} {J. Chem. Theory Comput.}\ }\textbf {\bibinfo {volume}
  {18}},\ \bibinfo {pages} {4945} (\bibinfo {year} {2022})}\BibitemShut
  {NoStop}%
\bibitem [{\citenamefont {Ma}\ \emph {et~al.}(2025)\citenamefont {Ma},
  \citenamefont {Dugan},\ and\ \citenamefont {Solja{\v{c}}i{\'c}}}]{Ma2025}%
  \BibitemOpen
  \bibfield  {author} {\bibinfo {author} {\bibfnamefont {A.}~\bibnamefont
  {Ma}}, \bibinfo {author} {\bibfnamefont {O.}~\bibnamefont {Dugan}},\ and\
  \bibinfo {author} {\bibfnamefont {M.}~\bibnamefont {Solja{\v{c}}i{\'c}}},\
  }\bibfield  {title} {\bibinfo {title} {{P}redicting band gap from chemical
  composition: {A} simple learned model for a material property with atypical
  statistics},\ }\href@noop {} {\bibfield  {journal} {\bibinfo  {journal}
  {arXiv preprint arXiv:2501.02932}\ } (\bibinfo {year} {2025})}\BibitemShut
  {NoStop}%
\bibitem [{\citenamefont {Saputro}\ and\ \citenamefont
  {Widyaningsih}(2017)}]{saputro2017}%
  \BibitemOpen
  \bibfield  {author} {\bibinfo {author} {\bibfnamefont {D.~R.~S.}\
  \bibnamefont {Saputro}}\ and\ \bibinfo {author} {\bibfnamefont
  {P.}~\bibnamefont {Widyaningsih}},\ }\bibfield  {title} {\bibinfo {title}
  {{L}imited memory {B}royden-{F}letcher-{G}oldfarb-{S}hanno ({L}-{B}{F}{G}{S})
  method for the parameter estimation on geographically weighted ordinal
  logistic regression model ({G}{W}{O}{L}{R})},\ }in\ \href@noop {} {\emph
  {\bibinfo {booktitle} {AIP conference proceedings}}},\ Vol.\ \bibinfo
  {volume} {1868}\ (\bibinfo {organization} {AIP Publishing LLC},\ \bibinfo
  {year} {2017})\ p.\ \bibinfo {pages} {040009}\BibitemShut {NoStop}%
\bibitem [{\citenamefont {Huang}\ \emph {et~al.}(2025)\citenamefont {Huang},
  \citenamefont {Huang}, \citenamefont {Xiao}, \citenamefont {Ma},
  \citenamefont {Ming}, \citenamefont {Shi},\ and\ \citenamefont
  {Wen}}]{huang2025}%
  \BibitemOpen
  \bibfield  {author} {\bibinfo {author} {\bibfnamefont {Z.}~\bibnamefont
  {Huang}}, \bibinfo {author} {\bibfnamefont {D.~Z.}\ \bibnamefont {Huang}},
  \bibinfo {author} {\bibfnamefont {T.}~\bibnamefont {Xiao}}, \bibinfo {author}
  {\bibfnamefont {D.}~\bibnamefont {Ma}}, \bibinfo {author} {\bibfnamefont
  {Z.}~\bibnamefont {Ming}}, \bibinfo {author} {\bibfnamefont {H.}~\bibnamefont
  {Shi}},\ and\ \bibinfo {author} {\bibfnamefont {Y.}~\bibnamefont {Wen}},\
  }\bibfield  {title} {\bibinfo {title} {{I}mproving {M}onte {C}arlo {T}ree
  {S}earch for {S}ymbolic {R}egression},\ }\href@noop {} {\bibfield  {journal}
  {\bibinfo  {journal} {arXiv preprint arXiv:2509.15929}\ } (\bibinfo {year}
  {2025})}\BibitemShut {NoStop}%
\bibitem [{\citenamefont {Xu}\ \emph {et~al.}(2023)\citenamefont {Xu},
  \citenamefont {Liu},\ and\ \citenamefont {Sun}}]{xu2023}%
  \BibitemOpen
  \bibfield  {author} {\bibinfo {author} {\bibfnamefont {Y.}~\bibnamefont
  {Xu}}, \bibinfo {author} {\bibfnamefont {Y.}~\bibnamefont {Liu}},\ and\
  \bibinfo {author} {\bibfnamefont {H.}~\bibnamefont {Sun}},\ }\bibfield
  {title} {\bibinfo {title} {{R}srm: {R}einforcement symbolic regression
  machine},\ }\href@noop {} {\bibfield  {journal} {\bibinfo  {journal} {arXiv
  preprint arXiv:2305.14656}\ } (\bibinfo {year} {2023})}\BibitemShut {NoStop}%
\bibitem [{\citenamefont {Landajuela}\ \emph {et~al.}(2022)\citenamefont
  {Landajuela}, \citenamefont {Lee}, \citenamefont {Yang}, \citenamefont
  {Glatt}, \citenamefont {Santiago}, \citenamefont {Aravena}, \citenamefont
  {Mundhenk}, \citenamefont {Mulcahy},\ and\ \citenamefont
  {Petersen}}]{landajuela2022}%
  \BibitemOpen
  \bibfield  {author} {\bibinfo {author} {\bibfnamefont {M.}~\bibnamefont
  {Landajuela}}, \bibinfo {author} {\bibfnamefont {C.~S.}\ \bibnamefont {Lee}},
  \bibinfo {author} {\bibfnamefont {J.}~\bibnamefont {Yang}}, \bibinfo {author}
  {\bibfnamefont {R.}~\bibnamefont {Glatt}}, \bibinfo {author} {\bibfnamefont
  {C.~P.}\ \bibnamefont {Santiago}}, \bibinfo {author} {\bibfnamefont
  {I.}~\bibnamefont {Aravena}}, \bibinfo {author} {\bibfnamefont
  {T.}~\bibnamefont {Mundhenk}}, \bibinfo {author} {\bibfnamefont
  {G.}~\bibnamefont {Mulcahy}},\ and\ \bibinfo {author} {\bibfnamefont {B.~K.}\
  \bibnamefont {Petersen}},\ }\bibfield  {title} {\bibinfo {title} {{A} unified
  framework for deep symbolic regression},\ }\href@noop {} {\bibfield
  {journal} {\bibinfo  {journal} {Adv. Neural Inf. Process. Syst.}\ }\textbf
  {\bibinfo {volume} {35}},\ \bibinfo {pages} {33985} (\bibinfo {year}
  {2022})}\BibitemShut {NoStop}%
\bibitem [{\citenamefont {Mundhenk}\ \emph {et~al.}(2021)\citenamefont
  {Mundhenk}, \citenamefont {Landajuela}, \citenamefont {Glatt}, \citenamefont
  {Santiago}, \citenamefont {Faissol},\ and\ \citenamefont
  {Petersen}}]{mundhenk2021}%
  \BibitemOpen
  \bibfield  {author} {\bibinfo {author} {\bibfnamefont {T.~N.}\ \bibnamefont
  {Mundhenk}}, \bibinfo {author} {\bibfnamefont {M.}~\bibnamefont
  {Landajuela}}, \bibinfo {author} {\bibfnamefont {R.}~\bibnamefont {Glatt}},
  \bibinfo {author} {\bibfnamefont {C.~P.}\ \bibnamefont {Santiago}}, \bibinfo
  {author} {\bibfnamefont {D.~M.}\ \bibnamefont {Faissol}},\ and\ \bibinfo
  {author} {\bibfnamefont {B.~K.}\ \bibnamefont {Petersen}},\ }\bibfield
  {title} {\bibinfo {title} {{S}ymbolic regression via neural-guided genetic
  programming population seeding},\ }\href@noop {} {\bibfield  {journal}
  {\bibinfo  {journal} {arXiv preprint arXiv:2111.00053}\ } (\bibinfo {year}
  {2021})}\BibitemShut {NoStop}%
\bibitem [{\citenamefont {Dunn}\ \emph {et~al.}({\natexlab{a}})\citenamefont
  {Dunn}, \citenamefont {Wang}, \citenamefont {Ganose}, \citenamefont {Dopp},\
  and\ \citenamefont {Jain}}]{matbench_expt_gap_leaderboard}%
  \BibitemOpen
  \bibfield  {author} {\bibinfo {author} {\bibfnamefont {A.}~\bibnamefont
  {Dunn}}, \bibinfo {author} {\bibfnamefont {Q.}~\bibnamefont {Wang}}, \bibinfo
  {author} {\bibfnamefont {A.}~\bibnamefont {Ganose}}, \bibinfo {author}
  {\bibfnamefont {D.}~\bibnamefont {Dopp}},\ and\ \bibinfo {author}
  {\bibfnamefont {A.}~\bibnamefont {Jain}},\ }\href@noop {} {\bibinfo {title}
  {{M}at{B}ench v0.1 {L}eaderboard: matbench\_expt\_gap}},\ \bibinfo
  {howpublished}
  {\url{https://matbench.materialsproject.org/Leaderboards\%20Per-Task/matbench_v0.1_matbench_expt_gap/}}
  ({\natexlab{a}}),\ \bibinfo {note} {accessed: 2026-05-04}\BibitemShut
  {NoStop}%
\bibitem [{\citenamefont {Dunn}\ \emph {et~al.}({\natexlab{b}})\citenamefont
  {Dunn}, \citenamefont {Wang}, \citenamefont {Ganose}, \citenamefont {Dopp},\
  and\ \citenamefont {Jain}}]{matbench_expt_is_metal_leaderboard}%
  \BibitemOpen
  \bibfield  {author} {\bibinfo {author} {\bibfnamefont {A.}~\bibnamefont
  {Dunn}}, \bibinfo {author} {\bibfnamefont {Q.}~\bibnamefont {Wang}}, \bibinfo
  {author} {\bibfnamefont {A.}~\bibnamefont {Ganose}}, \bibinfo {author}
  {\bibfnamefont {D.}~\bibnamefont {Dopp}},\ and\ \bibinfo {author}
  {\bibfnamefont {A.}~\bibnamefont {Jain}},\ }\href@noop {} {\bibinfo {title}
  {{M}atbench v0.1 {L}eaderboard: matbench\_expt\_is\_metal}},\ \bibinfo
  {howpublished}
  {\url{https://matbench.materialsproject.org/Leaderboards\%20Per-Task/matbench_v0.1_matbench_expt_is_metal/}}
  ({\natexlab{b}}),\ \bibinfo {note} {materials Project MatBench leaderboard,
  accessed: 2026-05-04}\BibitemShut {NoStop}%
\bibitem [{\citenamefont {Dunn}\ \emph {et~al.}({\natexlab{c}})\citenamefont
  {Dunn}, \citenamefont {Wang}, \citenamefont {Ganose}, \citenamefont {Dopp},\
  and\ \citenamefont {Jain}}]{matbench_glass_leaderboard}%
  \BibitemOpen
  \bibfield  {author} {\bibinfo {author} {\bibfnamefont {A.}~\bibnamefont
  {Dunn}}, \bibinfo {author} {\bibfnamefont {Q.}~\bibnamefont {Wang}}, \bibinfo
  {author} {\bibfnamefont {A.}~\bibnamefont {Ganose}}, \bibinfo {author}
  {\bibfnamefont {D.}~\bibnamefont {Dopp}},\ and\ \bibinfo {author}
  {\bibfnamefont {A.}~\bibnamefont {Jain}},\ }\href@noop {} {\bibinfo {title}
  {{M}atbench v0.1 {L}eaderboard: matbench\_glass}},\ \bibinfo {howpublished}
  {\url{https://matbench.materialsproject.org/Leaderboards\%20Per-Task/matbench_v0.1_matbench_glass/}}
  ({\natexlab{c}}),\ \bibinfo {note} {materials Project MatBench leaderboard,
  accessed: 2026-05-04}\BibitemShut {NoStop}%
\bibitem [{\citenamefont {Xie}\ \emph {et~al.}(2023)\citenamefont {Xie},
  \citenamefont {Wan}, \citenamefont {Huang}, \citenamefont {Yin},
  \citenamefont {Liu}, \citenamefont {Wang}, \citenamefont {Linghu},
  \citenamefont {Kit}, \citenamefont {Grazian}, \citenamefont {Zhang},
  \citenamefont {Razzak},\ and\ \citenamefont {Hoex}}]{xie2023}%
  \BibitemOpen
  \bibfield  {author} {\bibinfo {author} {\bibfnamefont {T.}~\bibnamefont
  {Xie}}, \bibinfo {author} {\bibfnamefont {Y.}~\bibnamefont {Wan}}, \bibinfo
  {author} {\bibfnamefont {W.}~\bibnamefont {Huang}}, \bibinfo {author}
  {\bibfnamefont {Z.}~\bibnamefont {Yin}}, \bibinfo {author} {\bibfnamefont
  {Y.}~\bibnamefont {Liu}}, \bibinfo {author} {\bibfnamefont {S.}~\bibnamefont
  {Wang}}, \bibinfo {author} {\bibfnamefont {Q.}~\bibnamefont {Linghu}},
  \bibinfo {author} {\bibfnamefont {C.}~\bibnamefont {Kit}}, \bibinfo {author}
  {\bibfnamefont {C.}~\bibnamefont {Grazian}}, \bibinfo {author} {\bibfnamefont
  {W.}~\bibnamefont {Zhang}}, \bibinfo {author} {\bibfnamefont
  {I.}~\bibnamefont {Razzak}},\ and\ \bibinfo {author} {\bibfnamefont
  {B.}~\bibnamefont {Hoex}},\ }\href@noop {} {\bibinfo {title}
  {{D}{A}{R}{W}{I}{N} {S}eries: {D}omain {S}pecific {L}arge {L}anguage {M}odels
  for {N}atural {S}cience}} (\bibinfo {year} {2023}),\ \Eprint
  {https://arxiv.org/abs/2308.13565} {arXiv:2308.13565 [cs.CL]} \BibitemShut
  {NoStop}%
\bibitem [{\citenamefont {Xie}\ \emph {et~al.}(2025)\citenamefont {Xie},
  \citenamefont {Wan}, \citenamefont {Liu}, \citenamefont {Zeng}, \citenamefont
  {Wang}, \citenamefont {Zhang}, \citenamefont {Grazian}, \citenamefont {Kit},
  \citenamefont {Ouyang}, \citenamefont {Zhou},\ and\ \citenamefont
  {Hoex}}]{xie2025}%
  \BibitemOpen
  \bibfield  {author} {\bibinfo {author} {\bibfnamefont {T.}~\bibnamefont
  {Xie}}, \bibinfo {author} {\bibfnamefont {Y.}~\bibnamefont {Wan}}, \bibinfo
  {author} {\bibfnamefont {Y.}~\bibnamefont {Liu}}, \bibinfo {author}
  {\bibfnamefont {Y.}~\bibnamefont {Zeng}}, \bibinfo {author} {\bibfnamefont
  {S.}~\bibnamefont {Wang}}, \bibinfo {author} {\bibfnamefont {W.}~\bibnamefont
  {Zhang}}, \bibinfo {author} {\bibfnamefont {C.}~\bibnamefont {Grazian}},
  \bibinfo {author} {\bibfnamefont {C.}~\bibnamefont {Kit}}, \bibinfo {author}
  {\bibfnamefont {W.}~\bibnamefont {Ouyang}}, \bibinfo {author} {\bibfnamefont
  {D.}~\bibnamefont {Zhou}},\ and\ \bibinfo {author} {\bibfnamefont
  {B.}~\bibnamefont {Hoex}},\ }\href {https://arxiv.org/abs/2412.11970}
  {\bibinfo {title} {{D}{A}{R}{W}{I}{N} 1.5: {L}arge {L}anguage {M}odels as
  {M}aterials {S}cience {A}dapted {L}earners}} (\bibinfo {year} {2025}),\
  \Eprint {https://arxiv.org/abs/2412.11970} {arXiv:2412.11970 [cs.CL]}
  \BibitemShut {NoStop}%
\bibitem [{\citenamefont {Jablonka}\ \emph {et~al.}(2024)\citenamefont
  {Jablonka}, \citenamefont {Schwaller}, \citenamefont {Ortega-Guerrero},\ and\
  \citenamefont {Smit}}]{jablonka2024}%
  \BibitemOpen
  \bibfield  {author} {\bibinfo {author} {\bibfnamefont {K.~M.}\ \bibnamefont
  {Jablonka}}, \bibinfo {author} {\bibfnamefont {P.}~\bibnamefont {Schwaller}},
  \bibinfo {author} {\bibfnamefont {A.}~\bibnamefont {Ortega-Guerrero}},\ and\
  \bibinfo {author} {\bibfnamefont {B.}~\bibnamefont {Smit}},\ }\bibfield
  {title} {\bibinfo {title} {{L}everaging large language models for predictive
  chemistry},\ }\href@noop {} {\bibfield  {journal} {\bibinfo  {journal} {Nat.
  Mach. Intell.}\ }\textbf {\bibinfo {volume} {6}},\ \bibinfo {pages} {161}
  (\bibinfo {year} {2024})}\BibitemShut {NoStop}%
\bibitem [{\citenamefont {Adachi}(2009)}]{adachi2009}%
  \BibitemOpen
  \bibfield  {author} {\bibinfo {author} {\bibfnamefont {S.}~\bibnamefont
  {Adachi}},\ }\href@noop {} {\emph {\bibinfo {title} {{P}roperties of
  semiconductor alloys: group-{I}{V}, {I}{I}{I}-{V} and {I}{I}-{V}{I}
  semiconductors}}}\ (\bibinfo  {publisher} {John Wiley \& Sons},\ \bibinfo
  {year} {2009})\BibitemShut {NoStop}%
\bibitem [{\citenamefont {Aubel}\ \emph {et~al.}(1985)\citenamefont {Aubel},
  \citenamefont {Reddy}, \citenamefont {Sundaram}, \citenamefont {Beard},\ and\
  \citenamefont {Comas}}]{aubel1985}%
  \BibitemOpen
  \bibfield  {author} {\bibinfo {author} {\bibfnamefont {J.}~\bibnamefont
  {Aubel}}, \bibinfo {author} {\bibfnamefont {U.}~\bibnamefont {Reddy}},
  \bibinfo {author} {\bibfnamefont {S.}~\bibnamefont {Sundaram}}, \bibinfo
  {author} {\bibfnamefont {W.}~\bibnamefont {Beard}},\ and\ \bibinfo {author}
  {\bibfnamefont {J.}~\bibnamefont {Comas}},\ }\bibfield  {title} {\bibinfo
  {title} {{I}nterband transitions in molecular-beam-epitaxial {A}l x {G}a1- x
  {A}s/{G}a{A}s},\ }\href@noop {} {\bibfield  {journal} {\bibinfo  {journal}
  {J. Appl. Phys.}\ }\textbf {\bibinfo {volume} {58}},\ \bibinfo {pages} {495}
  (\bibinfo {year} {1985})}\BibitemShut {NoStop}%
\bibitem [{\citenamefont {Aspnes}\ \emph {et~al.}(1986)\citenamefont {Aspnes},
  \citenamefont {Kelso}, \citenamefont {Logan},\ and\ \citenamefont
  {Bhat}}]{aspnes1986}%
  \BibitemOpen
  \bibfield  {author} {\bibinfo {author} {\bibfnamefont {D.}~\bibnamefont
  {Aspnes}}, \bibinfo {author} {\bibfnamefont {S.}~\bibnamefont {Kelso}},
  \bibinfo {author} {\bibfnamefont {R.}~\bibnamefont {Logan}},\ and\ \bibinfo
  {author} {\bibfnamefont {R.}~\bibnamefont {Bhat}},\ }\bibfield  {title}
  {\bibinfo {title} {{O}ptical properties of {A}l x {G}a1- x {A}s},\
  }\href@noop {} {\bibfield  {journal} {\bibinfo  {journal} {J. Appl. Phys.}\
  }\textbf {\bibinfo {volume} {60}},\ \bibinfo {pages} {754} (\bibinfo {year}
  {1986})}\BibitemShut {NoStop}%
\bibitem [{\citenamefont {Saxena}(1981)}]{saxena1981}%
  \BibitemOpen
  \bibfield  {author} {\bibinfo {author} {\bibfnamefont {A.}~\bibnamefont
  {Saxena}},\ }\bibfield  {title} {\bibinfo {title} {{N}on-$\gamma$ {D}eep
  {L}evels and the {C}onduction {B}and {S}tructure of {G}a1- x{A}lx{A}s
  {A}lloys},\ }\href@noop {} {\bibfield  {journal} {\bibinfo  {journal} {Phys.
  Status Solidi B}\ }\textbf {\bibinfo {volume} {105}},\ \bibinfo {pages} {777}
  (\bibinfo {year} {1981})}\BibitemShut {NoStop}%
\bibitem [{\citenamefont {Monemar}\ \emph {et~al.}(1976)\citenamefont
  {Monemar}, \citenamefont {Shih},\ and\ \citenamefont {Pettit}}]{monemar1976}%
  \BibitemOpen
  \bibfield  {author} {\bibinfo {author} {\bibfnamefont {B.}~\bibnamefont
  {Monemar}}, \bibinfo {author} {\bibfnamefont {K.}~\bibnamefont {Shih}},\ and\
  \bibinfo {author} {\bibfnamefont {G.}~\bibnamefont {Pettit}},\ }\bibfield
  {title} {\bibinfo {title} {{S}ome optical properties of the {A}l x {G}a1- x
  {A}s alloys system},\ }\href@noop {} {\bibfield  {journal} {\bibinfo
  {journal} {J. Appl. Phys.}\ }\textbf {\bibinfo {volume} {47}},\ \bibinfo
  {pages} {2604} (\bibinfo {year} {1976})}\BibitemShut {NoStop}%
\bibitem [{\citenamefont {Kim}\ \emph {et~al.}(2003)\citenamefont {Kim},
  \citenamefont {Ghong}, \citenamefont {Kim}, \citenamefont {Kim},
  \citenamefont {Aspnes}, \citenamefont {Mori}, \citenamefont {Yao},\ and\
  \citenamefont {Koo}}]{kim2003}%
  \BibitemOpen
  \bibfield  {author} {\bibinfo {author} {\bibfnamefont {T.}~\bibnamefont
  {Kim}}, \bibinfo {author} {\bibfnamefont {T.}~\bibnamefont {Ghong}}, \bibinfo
  {author} {\bibfnamefont {Y.}~\bibnamefont {Kim}}, \bibinfo {author}
  {\bibfnamefont {S.}~\bibnamefont {Kim}}, \bibinfo {author} {\bibfnamefont
  {D.}~\bibnamefont {Aspnes}}, \bibinfo {author} {\bibfnamefont
  {T.}~\bibnamefont {Mori}}, \bibinfo {author} {\bibfnamefont {T.}~\bibnamefont
  {Yao}},\ and\ \bibinfo {author} {\bibfnamefont {B.}~\bibnamefont {Koo}},\
  }\bibfield  {title} {\bibinfo {title} {{D}ielectric functions of {I}n x {G}a
  1- x {A}s alloys},\ }\href@noop {} {\bibfield  {journal} {\bibinfo  {journal}
  {Phys. Rev. B}\ }\textbf {\bibinfo {volume} {68}},\ \bibinfo {pages} {115323}
  (\bibinfo {year} {2003})}\BibitemShut {NoStop}%
\bibitem [{\citenamefont {Gaskill}\ \emph {et~al.}(1990)\citenamefont
  {Gaskill}, \citenamefont {Bottka}, \citenamefont {Aina},\ and\ \citenamefont
  {Mattingly}}]{gaskill1990}%
  \BibitemOpen
  \bibfield  {author} {\bibinfo {author} {\bibfnamefont {D.}~\bibnamefont
  {Gaskill}}, \bibinfo {author} {\bibfnamefont {N.}~\bibnamefont {Bottka}},
  \bibinfo {author} {\bibfnamefont {L.}~\bibnamefont {Aina}},\ and\ \bibinfo
  {author} {\bibfnamefont {M.}~\bibnamefont {Mattingly}},\ }\bibfield  {title}
  {\bibinfo {title} {{B}and-gap determination by photoreflectance of
  {I}n{G}a{A}s and {I}n{A}l{A}s lattice matched to {I}n{P}},\ }\href@noop {}
  {\bibfield  {journal} {\bibinfo  {journal} {Appl. Phys. Lett.}\ }\textbf
  {\bibinfo {volume} {56}},\ \bibinfo {pages} {1269} (\bibinfo {year}
  {1990})}\BibitemShut {NoStop}%
\bibitem [{\citenamefont {Woolley}\ and\ \citenamefont
  {Warner}(1964)}]{woolley1964}%
  \BibitemOpen
  \bibfield  {author} {\bibinfo {author} {\bibfnamefont {J.}~\bibnamefont
  {Woolley}}\ and\ \bibinfo {author} {\bibfnamefont {J.}~\bibnamefont
  {Warner}},\ }\bibfield  {title} {\bibinfo {title} {{O}ptical energy-gap
  variation in {I}n{A}s--{I}n{S}b alloys},\ }\href@noop {} {\bibfield
  {journal} {\bibinfo  {journal} {Can. J. Phys.}\ }\textbf {\bibinfo {volume}
  {42}},\ \bibinfo {pages} {1879} (\bibinfo {year} {1964})}\BibitemShut
  {NoStop}%
\bibitem [{\citenamefont {Dobbelaere}\ \emph {et~al.}(1989)\citenamefont
  {Dobbelaere}, \citenamefont {De~Boeck},\ and\ \citenamefont
  {Borghs}}]{dobbelaere1989}%
  \BibitemOpen
  \bibfield  {author} {\bibinfo {author} {\bibfnamefont {W.}~\bibnamefont
  {Dobbelaere}}, \bibinfo {author} {\bibfnamefont {J.}~\bibnamefont
  {De~Boeck}},\ and\ \bibinfo {author} {\bibfnamefont {G.}~\bibnamefont
  {Borghs}},\ }\bibfield  {title} {\bibinfo {title} {{G}rowth and optical
  characterization of {I}n{A}s1- x {S}b x (0$\le$ x$\le$ 1) on {G}a{A}s and on
  {G}a{A}s-coated {S}i by molecular beam epitaxy},\ }\href@noop {} {\bibfield
  {journal} {\bibinfo  {journal} {Appl. Phys. Lett.}\ }\textbf {\bibinfo
  {volume} {55}},\ \bibinfo {pages} {1856} (\bibinfo {year}
  {1989})}\BibitemShut {NoStop}%
\bibitem [{\citenamefont {Vishnubhatla}\ \emph {et~al.}(1969)\citenamefont
  {Vishnubhatla}, \citenamefont {Eyglunent},\ and\ \citenamefont
  {Woolley}}]{vishnubhatla1969}%
  \BibitemOpen
  \bibfield  {author} {\bibinfo {author} {\bibfnamefont {S.~S.}\ \bibnamefont
  {Vishnubhatla}}, \bibinfo {author} {\bibfnamefont {B.}~\bibnamefont
  {Eyglunent}},\ and\ \bibinfo {author} {\bibfnamefont {J.~C.}\ \bibnamefont
  {Woolley}},\ }\bibfield  {title} {\bibinfo {title} {{E}lectroreflectance
  measurements in mixed {I}{I}{I}--{V} alloys},\ }\href@noop {} {\bibfield
  {journal} {\bibinfo  {journal} {Can. J. Phys.}\ }\textbf {\bibinfo {volume}
  {47}},\ \bibinfo {pages} {1661} (\bibinfo {year} {1969})}\BibitemShut
  {NoStop}%
\bibitem [{\citenamefont {Auvergne}\ \emph {et~al.}(1974)\citenamefont
  {Auvergne}, \citenamefont {Camassel}, \citenamefont {Mathieu},\ and\
  \citenamefont {Joullie}}]{auvergne1974}%
  \BibitemOpen
  \bibfield  {author} {\bibinfo {author} {\bibfnamefont {D.}~\bibnamefont
  {Auvergne}}, \bibinfo {author} {\bibfnamefont {J.}~\bibnamefont {Camassel}},
  \bibinfo {author} {\bibfnamefont {H.}~\bibnamefont {Mathieu}},\ and\ \bibinfo
  {author} {\bibfnamefont {A.}~\bibnamefont {Joullie}},\ }\bibfield  {title}
  {\bibinfo {title} {{P}iezoreflectance measurements on {G}ax{I}n1- x {S}b
  alloys},\ }\href@noop {} {\bibfield  {journal} {\bibinfo  {journal} {J. Phys.
  Chem. Sol.}\ }\textbf {\bibinfo {volume} {35}},\ \bibinfo {pages} {133}
  (\bibinfo {year} {1974})}\BibitemShut {NoStop}%
\bibitem [{\citenamefont {Roth}\ and\ \citenamefont {Fortin}(1978)}]{roth1978}%
  \BibitemOpen
  \bibfield  {author} {\bibinfo {author} {\bibfnamefont {A.}~\bibnamefont
  {Roth}}\ and\ \bibinfo {author} {\bibfnamefont {E.}~\bibnamefont {Fortin}},\
  }\bibfield  {title} {\bibinfo {title} {{I}nterband magneto-optical study of
  the {I}n1- x {G}a x {S}b alloy system},\ }\href@noop {} {\bibfield  {journal}
  {\bibinfo  {journal} {Can. J. Phys.}\ }\textbf {\bibinfo {volume} {56}},\
  \bibinfo {pages} {1468} (\bibinfo {year} {1978})}\BibitemShut {NoStop}%
\bibitem [{\citenamefont {Alibert}\ \emph {et~al.}(1983)\citenamefont
  {Alibert}, \citenamefont {Joullie}, \citenamefont {Joullie},\ and\
  \citenamefont {Ance}}]{alibert1983}%
  \BibitemOpen
  \bibfield  {author} {\bibinfo {author} {\bibfnamefont {C.}~\bibnamefont
  {Alibert}}, \bibinfo {author} {\bibfnamefont {A.}~\bibnamefont {Joullie}},
  \bibinfo {author} {\bibfnamefont {A.}~\bibnamefont {Joullie}},\ and\ \bibinfo
  {author} {\bibfnamefont {C.}~\bibnamefont {Ance}},\ }\bibfield  {title}
  {\bibinfo {title} {{M}odulation-spectroscopy study of the {G}a 1- x {A}l x
  {S}b band structure},\ }\href@noop {} {\bibfield  {journal} {\bibinfo
  {journal} {Phys. Rev. B}\ }\textbf {\bibinfo {volume} {27}},\ \bibinfo
  {pages} {4946} (\bibinfo {year} {1983})}\BibitemShut {NoStop}%
\bibitem [{\citenamefont {Bignazzi}\ \emph {et~al.}(1998)\citenamefont
  {Bignazzi}, \citenamefont {Grilli}, \citenamefont {Guzzi}, \citenamefont
  {Bocchi}, \citenamefont {Bosacchi}, \citenamefont {Franchi},\ and\
  \citenamefont {Magnanini}}]{bignazzi1998}%
  \BibitemOpen
  \bibfield  {author} {\bibinfo {author} {\bibfnamefont {A.}~\bibnamefont
  {Bignazzi}}, \bibinfo {author} {\bibfnamefont {E.}~\bibnamefont {Grilli}},
  \bibinfo {author} {\bibfnamefont {M.}~\bibnamefont {Guzzi}}, \bibinfo
  {author} {\bibfnamefont {C.}~\bibnamefont {Bocchi}}, \bibinfo {author}
  {\bibfnamefont {A.}~\bibnamefont {Bosacchi}}, \bibinfo {author}
  {\bibfnamefont {S.}~\bibnamefont {Franchi}},\ and\ \bibinfo {author}
  {\bibfnamefont {R.}~\bibnamefont {Magnanini}},\ }\bibfield  {title} {\bibinfo
  {title} {{D}irect-and indirect-energy-gap dependence on {A}l concentration in
  {A}l x {G}a 1- x {S}b (x<\~{} 0. 4 1)},\ }\href@noop {} {\bibfield  {journal}
  {\bibinfo  {journal} {Phys. Rev. B}\ }\textbf {\bibinfo {volume} {57}},\
  \bibinfo {pages} {2295} (\bibinfo {year} {1998})}\BibitemShut {NoStop}%
\bibitem [{\citenamefont {Bellani}\ \emph {et~al.}(1999)\citenamefont
  {Bellani}, \citenamefont {Geddo}, \citenamefont {Guizzetti}, \citenamefont
  {Franchi},\ and\ \citenamefont {Magnanini}}]{bellani1999}%
  \BibitemOpen
  \bibfield  {author} {\bibinfo {author} {\bibfnamefont {V.}~\bibnamefont
  {Bellani}}, \bibinfo {author} {\bibfnamefont {M.}~\bibnamefont {Geddo}},
  \bibinfo {author} {\bibfnamefont {G.}~\bibnamefont {Guizzetti}}, \bibinfo
  {author} {\bibfnamefont {S.}~\bibnamefont {Franchi}},\ and\ \bibinfo {author}
  {\bibfnamefont {R.}~\bibnamefont {Magnanini}},\ }\bibfield  {title} {\bibinfo
  {title} {{T}hermoreflectance study of the direct optical gap in epitaxial
  {A}l x {G}a 1- x {S}b (x<\~{} 0. 5)},\ }\href@noop {} {\bibfield  {journal}
  {\bibinfo  {journal} {Phys. Rev. B}\ }\textbf {\bibinfo {volume} {59}},\
  \bibinfo {pages} {12272} (\bibinfo {year} {1999})}\BibitemShut {NoStop}%
\bibitem [{\citenamefont {Saadallah}\ \emph {et~al.}(2003)\citenamefont
  {Saadallah}, \citenamefont {Yacoubi}, \citenamefont {Genty},\ and\
  \citenamefont {Alibert}}]{saadallah2003}%
  \BibitemOpen
  \bibfield  {author} {\bibinfo {author} {\bibfnamefont {F.}~\bibnamefont
  {Saadallah}}, \bibinfo {author} {\bibfnamefont {N.}~\bibnamefont {Yacoubi}},
  \bibinfo {author} {\bibfnamefont {F.}~\bibnamefont {Genty}},\ and\ \bibinfo
  {author} {\bibfnamefont {C.}~\bibnamefont {Alibert}},\ }\bibfield  {title}
  {\bibinfo {title} {{P}hotothermal investigations of thermal and optical
  properties of {G}a{A}l{A}s{S}b and {A}l{A}s{S}b thin layers},\ }\href@noop {}
  {\bibfield  {journal} {\bibinfo  {journal} {J. Appl. Phys.}\ }\textbf
  {\bibinfo {volume} {94}},\ \bibinfo {pages} {5041} (\bibinfo {year}
  {2003})}\BibitemShut {NoStop}%
\bibitem [{\citenamefont {Rodriguez}\ and\ \citenamefont
  {Armelles}(1991)}]{rodriguez1991}%
  \BibitemOpen
  \bibfield  {author} {\bibinfo {author} {\bibfnamefont {J.}~\bibnamefont
  {Rodriguez}}\ and\ \bibinfo {author} {\bibfnamefont {G.}~\bibnamefont
  {Armelles}},\ }\bibfield  {title} {\bibinfo {title} {{E}llipsometric study of
  {A}l{I}n{A}s and {A}l{G}a{P} alloys},\ }\href@noop {} {\bibfield  {journal}
  {\bibinfo  {journal} {J. Appl. Phys.}\ }\textbf {\bibinfo {volume} {69}},\
  \bibinfo {pages} {965} (\bibinfo {year} {1991})}\BibitemShut {NoStop}%
\bibitem [{\citenamefont {Choi}\ \emph {et~al.}(2000)\citenamefont {Choi},
  \citenamefont {Kim}, \citenamefont {Yoo}, \citenamefont {Aspnes},
  \citenamefont {Woo},\ and\ \citenamefont {Kim}}]{choi2000}%
  \BibitemOpen
  \bibfield  {author} {\bibinfo {author} {\bibfnamefont {S.}~\bibnamefont
  {Choi}}, \bibinfo {author} {\bibfnamefont {Y.}~\bibnamefont {Kim}}, \bibinfo
  {author} {\bibfnamefont {S.}~\bibnamefont {Yoo}}, \bibinfo {author}
  {\bibfnamefont {D.}~\bibnamefont {Aspnes}}, \bibinfo {author} {\bibfnamefont
  {D.}~\bibnamefont {Woo}},\ and\ \bibinfo {author} {\bibfnamefont
  {S.}~\bibnamefont {Kim}},\ }\bibfield  {title} {\bibinfo {title} {{O}ptical
  properties of {A}l x {G}a 1- x {P} (0$\le$ x$\le$ 0.52) alloys},\ }\href@noop
  {} {\bibfield  {journal} {\bibinfo  {journal} {J. Appl. Phys.}\ }\textbf
  {\bibinfo {volume} {87}},\ \bibinfo {pages} {1287} (\bibinfo {year}
  {2000})}\BibitemShut {NoStop}%
\bibitem [{\citenamefont {Onton}\ \emph {et~al.}(1971)\citenamefont {Onton},
  \citenamefont {Lorenz},\ and\ \citenamefont {Reuter}}]{onton1971}%
  \BibitemOpen
  \bibfield  {author} {\bibinfo {author} {\bibfnamefont {A.}~\bibnamefont
  {Onton}}, \bibinfo {author} {\bibfnamefont {M.}~\bibnamefont {Lorenz}},\ and\
  \bibinfo {author} {\bibfnamefont {W.}~\bibnamefont {Reuter}},\ }\bibfield
  {title} {\bibinfo {title} {{E}lectronic {S}tructure and {L}uminescence
  {P}rocesses in {I}n1- x {G}a x {P} {A}lloys},\ }\href@noop {} {\bibfield
  {journal} {\bibinfo  {journal} {J. Appl. Phys.}\ }\textbf {\bibinfo {volume}
  {42}},\ \bibinfo {pages} {3420} (\bibinfo {year} {1971})}\BibitemShut
  {NoStop}%
\bibitem [{\citenamefont {Alibert}\ \emph {et~al.}(1972)\citenamefont
  {Alibert}, \citenamefont {Bordure}, \citenamefont {Laugier},\ and\
  \citenamefont {Chevallier}}]{alibert1972}%
  \BibitemOpen
  \bibfield  {author} {\bibinfo {author} {\bibfnamefont {C.}~\bibnamefont
  {Alibert}}, \bibinfo {author} {\bibfnamefont {G.}~\bibnamefont {Bordure}},
  \bibinfo {author} {\bibfnamefont {A.}~\bibnamefont {Laugier}},\ and\ \bibinfo
  {author} {\bibfnamefont {J.}~\bibnamefont {Chevallier}},\ }\bibfield  {title}
  {\bibinfo {title} {{E}lectroreflectance and band structure of {G}a x {I}n 1-
  x {P} alloys},\ }\href@noop {} {\bibfield  {journal} {\bibinfo  {journal}
  {Phys. Rev. B}\ }\textbf {\bibinfo {volume} {6}},\ \bibinfo {pages} {1301}
  (\bibinfo {year} {1972})}\BibitemShut {NoStop}%
\bibitem [{\citenamefont {Sch{\"o}rmann}\ \emph {et~al.}(2006)\citenamefont
  {Sch{\"o}rmann}, \citenamefont {As}, \citenamefont {Lischka}, \citenamefont
  {Schley}, \citenamefont {Goldhahn}, \citenamefont {Li}, \citenamefont
  {L{\"o}ffler}, \citenamefont {Hetterich},\ and\ \citenamefont
  {Kalt}}]{schormann2006}%
  \BibitemOpen
  \bibfield  {author} {\bibinfo {author} {\bibfnamefont {J.}~\bibnamefont
  {Sch{\"o}rmann}}, \bibinfo {author} {\bibfnamefont {D.}~\bibnamefont {As}},
  \bibinfo {author} {\bibfnamefont {K.}~\bibnamefont {Lischka}}, \bibinfo
  {author} {\bibfnamefont {P.}~\bibnamefont {Schley}}, \bibinfo {author}
  {\bibfnamefont {R.}~\bibnamefont {Goldhahn}}, \bibinfo {author}
  {\bibfnamefont {S.}~\bibnamefont {Li}}, \bibinfo {author} {\bibfnamefont
  {W.}~\bibnamefont {L{\"o}ffler}}, \bibinfo {author} {\bibfnamefont
  {M.}~\bibnamefont {Hetterich}},\ and\ \bibinfo {author} {\bibfnamefont
  {H.}~\bibnamefont {Kalt}},\ }\bibfield  {title} {\bibinfo {title}
  {{M}olecular beam epitaxy of phase pure cubic {I}n{N}},\ }\href@noop {}
  {\bibfield  {journal} {\bibinfo  {journal} {Appl. Phys. Lett.}\ }\textbf
  {\bibinfo {volume} {89}} (\bibinfo {year} {2006})}\BibitemShut {NoStop}%
\bibitem [{\citenamefont {M{\"u}llh{\"a}user}\ \emph
  {et~al.}(1998)\citenamefont {M{\"u}llh{\"a}user}, \citenamefont {Brandt},
  \citenamefont {Trampert}, \citenamefont {Jenichen},\ and\ \citenamefont
  {Ploog}}]{mullhauser1998}%
  \BibitemOpen
  \bibfield  {author} {\bibinfo {author} {\bibfnamefont {J.}~\bibnamefont
  {M{\"u}llh{\"a}user}}, \bibinfo {author} {\bibfnamefont {O.}~\bibnamefont
  {Brandt}}, \bibinfo {author} {\bibfnamefont {A.}~\bibnamefont {Trampert}},
  \bibinfo {author} {\bibfnamefont {B.}~\bibnamefont {Jenichen}},\ and\
  \bibinfo {author} {\bibfnamefont {K.}~\bibnamefont {Ploog}},\ }\bibfield
  {title} {\bibinfo {title} {{G}reen photoluminescence from cubic {I}n 0.4 {G}a
  0.6 {N} grown by radio frequency plasma-assisted molecular beam epitaxy},\
  }\href@noop {} {\bibfield  {journal} {\bibinfo  {journal} {Appl. Phys.
  Lett.}\ }\textbf {\bibinfo {volume} {73}},\ \bibinfo {pages} {1230} (\bibinfo
  {year} {1998})}\BibitemShut {NoStop}%
\bibitem [{\citenamefont {Goldhahn}\ \emph {et~al.}(2000)\citenamefont
  {Goldhahn}, \citenamefont {Scheiner}, \citenamefont {Shokhovets},
  \citenamefont {Frey}, \citenamefont {K{\"o}hler}, \citenamefont {As},\ and\
  \citenamefont {Lischka}}]{goldhahn2000}%
  \BibitemOpen
  \bibfield  {author} {\bibinfo {author} {\bibfnamefont {R.}~\bibnamefont
  {Goldhahn}}, \bibinfo {author} {\bibfnamefont {J.}~\bibnamefont {Scheiner}},
  \bibinfo {author} {\bibfnamefont {S.}~\bibnamefont {Shokhovets}}, \bibinfo
  {author} {\bibfnamefont {T.}~\bibnamefont {Frey}}, \bibinfo {author}
  {\bibfnamefont {U.}~\bibnamefont {K{\"o}hler}}, \bibinfo {author}
  {\bibfnamefont {D.}~\bibnamefont {As}},\ and\ \bibinfo {author}
  {\bibfnamefont {K.}~\bibnamefont {Lischka}},\ }\bibfield  {title} {\bibinfo
  {title} {{R}efractive index and gap energy of cubic {I}n x {G}a 1- x {N}},\
  }\href@noop {} {\bibfield  {journal} {\bibinfo  {journal} {Appl. Phys.
  Lett.}\ }\textbf {\bibinfo {volume} {76}},\ \bibinfo {pages} {291} (\bibinfo
  {year} {2000})}\BibitemShut {NoStop}%
\bibitem [{\citenamefont {Takanobu~Suzuki}\ \emph {et~al.}(2000)\citenamefont
  {Takanobu~Suzuki}, \citenamefont {Hiroyuki~Yaguchi}, \citenamefont
  {Hajime~Okumura}, \citenamefont {Yuuki~Ishida},\ and\ \citenamefont
  {Sadafumi~Yoshida}}]{takanobu2000}%
  \BibitemOpen
  \bibfield  {author} {\bibinfo {author} {\bibfnamefont {T.~S.}\ \bibnamefont
  {Takanobu~Suzuki}}, \bibinfo {author} {\bibfnamefont {H.~Y.}\ \bibnamefont
  {Hiroyuki~Yaguchi}}, \bibinfo {author} {\bibfnamefont {H.~O.}\ \bibnamefont
  {Hajime~Okumura}}, \bibinfo {author} {\bibfnamefont {Y.~I.}\ \bibnamefont
  {Yuuki~Ishida}},\ and\ \bibinfo {author} {\bibfnamefont {S.~Y.}\ \bibnamefont
  {Sadafumi~Yoshida}},\ }\bibfield  {title} {\bibinfo {title} {{O}ptical
  constants of cubic {G}a{N}, {A}l{N}, and {A}l{G}a{N} alloys},\ }\href@noop {}
  {\bibfield  {journal} {\bibinfo  {journal} {Jpn. J. Appl. Phys.}\ }\textbf
  {\bibinfo {volume} {39}},\ \bibinfo {pages} {L497} (\bibinfo {year}
  {2000})}\BibitemShut {NoStop}%
\bibitem [{\citenamefont {Kasic}\ \emph {et~al.}(2002)\citenamefont {Kasic},
  \citenamefont {Schubert}, \citenamefont {Frey}, \citenamefont {K{\"o}hler},
  \citenamefont {As},\ and\ \citenamefont {Herzinger}}]{kasic2002}%
  \BibitemOpen
  \bibfield  {author} {\bibinfo {author} {\bibfnamefont {A.}~\bibnamefont
  {Kasic}}, \bibinfo {author} {\bibfnamefont {M.}~\bibnamefont {Schubert}},
  \bibinfo {author} {\bibfnamefont {T.}~\bibnamefont {Frey}}, \bibinfo {author}
  {\bibfnamefont {U.}~\bibnamefont {K{\"o}hler}}, \bibinfo {author}
  {\bibfnamefont {D.}~\bibnamefont {As}},\ and\ \bibinfo {author}
  {\bibfnamefont {C.}~\bibnamefont {Herzinger}},\ }\bibfield  {title} {\bibinfo
  {title} {{O}ptical phonon modes and interband transitions in cubic {A}l x
  {G}a 1- x {N} films},\ }\href@noop {} {\bibfield  {journal} {\bibinfo
  {journal} {Phys. Rev. B}\ }\textbf {\bibinfo {volume} {65}},\ \bibinfo
  {pages} {184302} (\bibinfo {year} {2002})}\BibitemShut {NoStop}%
\bibitem [{\citenamefont {Kearns}\ \emph {et~al.}(2002)\citenamefont {Kearns},
  \citenamefont {Mansour},\ and\ \citenamefont {Ng}}]{kearns2002}%
  \BibitemOpen
  \bibfield  {author} {\bibinfo {author} {\bibfnamefont {M.}~\bibnamefont
  {Kearns}}, \bibinfo {author} {\bibfnamefont {Y.}~\bibnamefont {Mansour}},\
  and\ \bibinfo {author} {\bibfnamefont {A.~Y.}\ \bibnamefont {Ng}},\
  }\bibfield  {title} {\bibinfo {title} {{A} sparse sampling algorithm for
  near-optimal planning in large {M}arkov decision processes},\ }\href@noop {}
  {\bibfield  {journal} {\bibinfo  {journal} {Mach. Learn.}\ }\textbf {\bibinfo
  {volume} {49}},\ \bibinfo {pages} {193} (\bibinfo {year} {2002})}\BibitemShut
  {NoStop}%
\bibitem [{\citenamefont {Kocsis}\ and\ \citenamefont
  {Szepesv{\'a}ri}(2006)}]{kocsis2006}%
  \BibitemOpen
  \bibfield  {author} {\bibinfo {author} {\bibfnamefont {L.}~\bibnamefont
  {Kocsis}}\ and\ \bibinfo {author} {\bibfnamefont {C.}~\bibnamefont
  {Szepesv{\'a}ri}},\ }\bibfield  {title} {\bibinfo {title} {{B}andit based
  monte-carlo planning},\ }in\ \href@noop {} {\emph {\bibinfo {booktitle}
  {European conference on machine learning}}}\ (\bibinfo {organization}
  {Springer},\ \bibinfo {year} {2006})\ pp.\ \bibinfo {pages}
  {282--293}\BibitemShut {NoStop}%
\end{thebibliography}%

\end{document}